\documentclass[apj]{emulateapj}
\usepackage{color}

\shorttitle{High-$z$ Low-luminosity QLF}
\shortauthors{Niida et al.}

\begin{document}

\title{REVISITING THE COMPLETENESS AND THE LUMINOSITY FUNCTION IN 
HIGH-REDSHIFT LOW-LUMINOSITY QUASAR SURVEYS}
    
\author{
{MANA~NIIDA}\altaffilmark{1},
{TOHRU~NAGAO}\altaffilmark{2}, 
HIROYUKI~IKEDA\altaffilmark{3}, 
KENTA~MATSUOKA\altaffilmark{4}, 
MASAKAZU~A.~R.~KOBAYASHI\altaffilmark{5}, 
YOSHIKI~TOBA\altaffilmark{2,6}, AND 
YOSHIAKI~TANIGUCHI\altaffilmark{7}
}

\altaffiltext{1}{Graduate School of Science and Engineering, Ehime University, 
Bunkyo-cho 2-5, Matsuyama 790-8577, Japan}
\email{niida@cosmos.phys.sci.ehime-u.ac.jp}
\altaffiltext{2}{Research Center for Space and Cosmic Evolution, Ehime
University, Bunkyo-cho 2-5, Matsuyama 790-8577, Japan}
\altaffiltext{3}{National Astronomical Observatory of Japan, Mitaka, Tokyo
181-8588, Japan}
\altaffiltext{4}{Department of Astronomy, Kyoto University, Kitashirakawa-Oiwake-cho, Sakyo-ku, Kyoto 606-8502, Japan}
\altaffiltext{5}{Faculty of Natural Sciences, National Institute of Technology, Kure College, 2-2-11, Agaminami, Kure, Hiroshima 737-8506, Japan}
\altaffiltext{6}{Academia Sinica Institute of Astronomy and Astrophysics, PO Box 23-141, Taipei 10617, Taiwan}
\altaffiltext{7}{The Open University of Japan, 2-11, Wakaba, Mihama-ku, Chiba 261-8586, Japan}

\begin{abstract}
Recent studies have derived quasar luminosity functions (QLFs) at various redshifts. However, the faint side 
of the QLF at high redshifts is still too uncertain. An accurate estimate of the survey completeness is 
essential to derive an accurate QLF for use in studying the luminosity-dependent density evolution of the quasar
population. Here we investigate how the luminosity dependence of quasar 
spectra (the Baldwin effect) and the attenuation model for the inter-galactic medium (IGM) affect the completeness 
estimates. For this purpose, we revisit the completeness of quasar surveys specifically at $z\sim4-5$, 
using the COSMOS images observed with Subaru/Suprime-Cam. As the result, we find that the completeness 
estimates are sensitive to the luminosity dependence of the quasar spectrum and difference in the 
IGM attenuation models. At $z\sim4$, the number density of quasars when we adopt the latest IGM model 
and take the luminosity dependence of spectra into account are $(3.49\pm1.62)\times10^{-7}$ Mpc$^{-3}$ mag$^{-1}$ 
for $-24.09<M_{1450}<-23.09$ and $(5.24\pm2.13)\times10^{-7}$ Mpc$^{-3}$ mag$^{-1}$ for $-23.09<M_{1450}<-22.09$,
respectively, which are $\sim$24\% lower than that estimated by the conventional method. On the other 
hand, at $z\sim5$, a $1\sigma$ confidence upper limit of the number density at $-24.5<M_{1450}<-22.5$ 
in our new estimates is $\sim$43\% higher than that estimated previously. The results suggest that 
the luminosity dependence of the quasar spectrum and the IGM model are important for deriving accurate number 
density of high-$z$ quasars. Even taking these effects into account, the inferred luminosity-dependent density evolution 
of quasars is consistent with the AGN down-sizing evolutionary picture.
\end{abstract}


\section{Introduction}\label{sec:1}

It is now widely recognized that the huge radiative energy released from active 
galactic nuclei (AGNs) is powered by the gravitational energy of supermassive 
black holes (SMBHs) at the center of AGNs \citep[e.g.,][]{ree84}. The mass 
of SMBHs ($M_{\rm BH}$) in quasars, that belong to the most luminous class of 
AGNs, reaches up to $\sim10^9 M_\odot$ or even more \citep[e.g.,][]{wil10, she12}.  
Interestingly, such massive SMBHs are seen even at very high redshifts, 
$z \sim 6-7$ \citep[e.g.,][]{kur07, mor11, ven13, ven15, wu15}. However, it is 
totally unclear when and how those SMBHs have formed and evolved. On the 
other hand, it has been observationally reported that there is a tight correlation 
between the mass of the host bulges ($M_{\rm bulge}$) and $M_{\rm BH}$ 
\citep[e.g.,][]{mahu03, hari04, gul09}. This correlation suggests that the SMBHs 
and their host galaxies have evolved with the close interplay, that is now 
recognized as the galaxy-SMBH coevolution. Therefore the observational study 
on the redshift evolution of SMBHs is important also for understanding the total 
picture of the galaxy evolution, not only for understanding the evolution of SMBHs 
themselves. 

For exploring the cosmological evolution of SMBHs observationally, one important 
class of AGNs is the quasar. This is because (i) quasars are in the phase where 
SMBHs are acquiring their mass via the gas accretion very actively and (ii) the 
huge luminosity of quasars enables us to measure $M_{\rm BH}$ through 
spectroscopic observations, even in the very distant Universe. Even without spectra, 
a very rough guess of $M_{\rm BH}$ can be obtained through their luminosity by 
assuming a certain Eddington ratio. In this sense, it is very interesting to 
investigate the number density evolution of quasars as a function of the redshift 
and luminosity; i.e., the redshift evolution of the quasar luminosity function (QLF). 
In other words, the accurate measurement of the QLF at a wide redshift range is 
a promising approach to study the cosmological evolution of SMBHs.

The QLF at $z \lesssim 3$ has been extensively measured over a wide luminosity 
range \citep[e.g.,][]{sia08, cro09, ros13, pal13}, and accordingly it is recognized that the QLF is 
expressed by the broken power-law formula.
Though the brighter part of the QLF at $z \gtrsim 4$ has been also measured 
\citep{ric06, she12}, the faint side of the QLF is still controversial 
\citep[e.g.,][]{gli10, gli11, ike11, ike12, mas12}.
Once focusing on high-luminosity quasars, previous observations suggest that 
the number density of luminous quasars increased from the early Universe to 
$z \sim 2$ and then decreased to the current Universe \citep[e.g.,][]{ric06, cro09}. 
It is more interesting to study possible luminosity dependences of the quasar 
number-density evolution (luminosity-dependent density evolution; LDDE). Recent 
optical surveys of high-redshift quasars have reported that lower-luminosity 
quasars show the peak of the number density at lower redshifts than 
higher-luminosity quasars \citep[e.g.,][]{cro09, ike11, ike12}. Since the quasar
luminosity at a given Eddington ratio corresponds to $M_{\rm BH}$,  the reported
LDDE trend is that is sometimes called the down-sizing evolution. Such a 
down-sizing trend has been reported also by some X-ray surveys 
(e.g., \citealt{ued03, ued14, has05, miy15}; see also \citealt{eno14} and 
references therein for theoretical works on the AGN down-sizing evolution). 
Note that the down-sizing evolution has been originally discussed for explaining 
the redshift evolution of the galaxy number density and mass function 
\citep[e.g.,][]{cow96, nei06, fon09}. Therefore the observational study on the 
AGN down-sizing evolution is exciting also for exploring the galaxy-SMBH 
co-evolution.

One caveat in the previous study on the AGN down-sizing evolution is that the
number density of low-luminosity quasars at high redshifts is quite uncertain, that 
prevents us from understanding the total picture of the quasar LDDE. \citet{ued14} 
mentioned that the number ratio of low-luminosity quasars to high-luminosity 
quasars is possibly higher at $z > 3$ (``up-down sizing''; see 
also \citealt{gli10, gli11}), but such detailed studies on the quasar LDDE at high 
redshifts required more statistically reliable QLFs with a wide luminosity range. 
Here we point out two issues that may introduce large systematic errors in the 
previous QLF studies. One is the effect of the luminosity dependence of quasar 
spectra on the completeness in quasar surveys. The survey 
completeness is usually estimated by adopting a typical template of the quasar 
spectrum, constructed empirically  \citep[e.g.,][]{van01} or based on 
simple models \citep[e.g.,][]{ike11, ike12}. However the equivalent width ($EW$) 
of broad emission lines in the quasar spectrum strongly depends on the quasar 
luminosity, in the sense that lower-luminosity quasars show emission lines with a 
larger $EW$ (the Baldwin effect; \citealt{bal77, kin90, bala04}; see also  
\citealt{nag06} and references therein for the luminosity dependence of flux ratios 
of quasar broad lines). If the quasar luminosity investigated in a survey is much 
lower than that used for the template in the completeness, the quasar 
colors calculated from the template may be systematically different from the 
actual quasar colors due to the unexpected contribution of emission-line fluxes 
into broad-band magnitudes. Therefore it is essential to know how the Baldwin 
effect affects the completeness in quasar surveys and the resultant QLFs. 
Another possible source of systematic errors in the QLF study is the inter-galactic medium (IGM) 
attenuation by neutral hydrogen, that is also important in the completeness 
estimate because the color of high-redshift quasars is sensitive to the IGM 
absorption. Recently \citet{ino14} pointed out that the previous models of the IGM 
attenuation (such as the model by \citealt{mad95}) tend to overestimate the 
optical depth of the IGM. Thus it is important to examine how the derived QLF is 
different from the previous estimate when we adopt the recent IGM model by  
\citet{ino14} that is more consistent with recent observational data of quasar 
absorption-line systems.

In this paper, we investigate how the Baldwin effect and the IGM models affect 
the quasar colors, completeness estimates, and the resultant QLFs, specifically 
focusing on quasars at $z \sim 4-5$. Based on the obtained results, we revisit 
the redshift evolution of the quasar space density in a wide luminosity range to 
examine the quasar LDDE. This paper is organized as follow. 
In Section~\ref{sec:2}, we describe how the quasar color is affected by the 
Baldwin effect and the adopted IGM model. In Section~\ref{sec:3} and 
Section~\ref{sec:4}, we show the resultant survey completeness and the QLF for 
$z \sim 4$ and 5. Finally we present the summary of this paper in 
Section~\ref{sec:5}. Throughout this paper, we adopt a $\Lambda$CDM cosmology 
with $\Omega_{\rm m} = 0.3$, $\Omega_\Lambda = 0.7$, and the Hubble 
constant of $H_0 = 70$ km s$^{-1}$ Mpc$^{-1}$.

\section{DATA}\label{sec:2}

\subsection{BOSS Quasars} 

For examining how the Baldwin effect affects the quasar colors, the $EW$ of 
emission lines should be known as a function of the quasar luminosity. To 
measure the $EW$, we focus on the archival data of the Baryon Oscillation
Spectroscopic Survey \citep[BOSS;][]{daw13} and investigate the spectra of 
quasars in the BOSS quasar catalog \citep{par12} in the Data Release 9 
\citep[DR9;][]{ahn12} of the Sloan Digital Sky Survey \citep[SDSS;][]{yor00}. 
In the BOSS observations, two double-armed spectrographs, which are 
upgraded and cover a wider spectral range than those used by the former 
SDSS observations, are used \citep{sme13}. Furthermore the BOSS targeted 
quasars with $\sim$ 2 magnitudes fainter than the original SDSS spectroscopic 
targets, i.e., $g^\ast < 22$ or $i^\ast < 21$ \citep{par12}. This wide magnitude 
range of the quasar sample is useful for our analysis of the Baldwin effect. The 
spectral resolution varies from $R$ $\sim$ 1,300 at 3600$\ \rm{\AA}$ to 
$\sim$ 2,500 at 10000$\ \rm{\AA}$ \citep{sme13}. This resolution is enough for 
our analysis, since the broad emission lines in quasars have a large velocity 
width typically ($>$ 1,000 km s$^{-1}$). 

The total area covered by the DR9 is 3,275 deg$^2$, in which 87,822 
spectroscopically confirmed quasars are identified \citep{par12}. In order to cover 
most strong rest-frame UV emission lines such as Ly$\alpha$ and {C~{\sc iv}}, 
we specifically focus on the redshift range of $2 \le z < 5$ where 65,419 quasars 
are listed in the catalog of \citet{par12}. Among them, quasars showing broad 
abosrption-line (BAL) features in their spectra (6,449 objects, that are identified 
through the \verb|BAL|$\_$\verb|FLAG| parameter in the catalog of \citealt{par12}) 
are removed because we cannot measure the $EW$ correctly in those cases. Then 
we select objects whose redshift is accurately measured, by adopting the criteria 
of \verb|ZWARNING| flag $= 0$ and \verb|ERR|\_\verb|ZPIPE| $\le 0.001$ (52,290 
objects). An additional criterion is also adopted for avoiding the mis-identification 
of emission lines (and consequently the wrong redshift), by comparing the redshift 
measured through the SDSS pipeline (\verb|Z|\_\verb|PIPE|) and that measured 
through the visual inspection (\verb|Z|\_\verb|VI|). In this step, 161 objects 
satisfying $|$\verb|Z|\_\verb|VI| -- \verb|Z|\_\verb|PIPE|$|$ $>$ 0.05 are removed. 
For selecting quasars with a spectrum which has a high signal-to-noise ratio, we 
adopt another criterion of \verb|SNR|\_\verb|SPEC| $> 1$ (43,962 objects). Then 
we selected 43,956 objects whose luminosity is in the range of $-31 \le M_i[z = 2] < -23$,
where $M_i[z = 2]$ is the $K$-corrected $i$-band magnitude at $z = 2$ \citep[see][]{ric06, ros13}, because in this luminosity range there 
are enough number of quasars for each luminosity. Finally, we remove three 
quasars whose FITS header has a problem. The number of finally selected BOSS
quasars in our analysis is 43,953.

\subsection{Composite Spectra of BOSS Quasars} \label{subsec:2.2}

To investigate the relation between the quasar luminosity and $EW$ of broad 
emission lines in the quasar spectrum statistically, we analyze composite spectra 
of BOSS quasars for each luminosity. This stacking analysis is a powerful method 
for studying the luminosity dependence of quasar spectral features, since it 
minimizes the dispersion of spectral features in quasar spectra in each luminosity 
bin. Another advantage of the stacking analysis is measuring the spectral features 
of low-luminosity quasars whose individual spectra show very faint spectral features 
with very low signal-to-noise ratios \citep[see, e.g.,][]{van01, nag06}.

First, we confirm that $EW_{\rm rest}$(C~{\sc iv}) of BOSS quasars does not depend 
on redshift, that is consistent with earlier works  \citep[e.g.,][see Appendix]{die02, cro02}. 
Therefore we combine the spectra of quasars in the whole 
redshift range of $2 \leq z < 5$ to make the composite spectra as a function of the 
absolute magnitude. This enables us to examine the luminosity dependence in a wide 
luminosity range, because quasars at higher redshift and those at lower redshift trace 
the higher luminosity and lower luminosity ranges, respectively. We divide the absolute 
magnitude range of $-31 \le M_i[z = 2] < -23$ into 8 magnitude bins with the bin width 
of $\Delta M_i[z = 2] = 1$ (see Table~\ref{tbl-1}), through the following analysis.

\begin{figure}
\epsscale{1.1}
\plotone{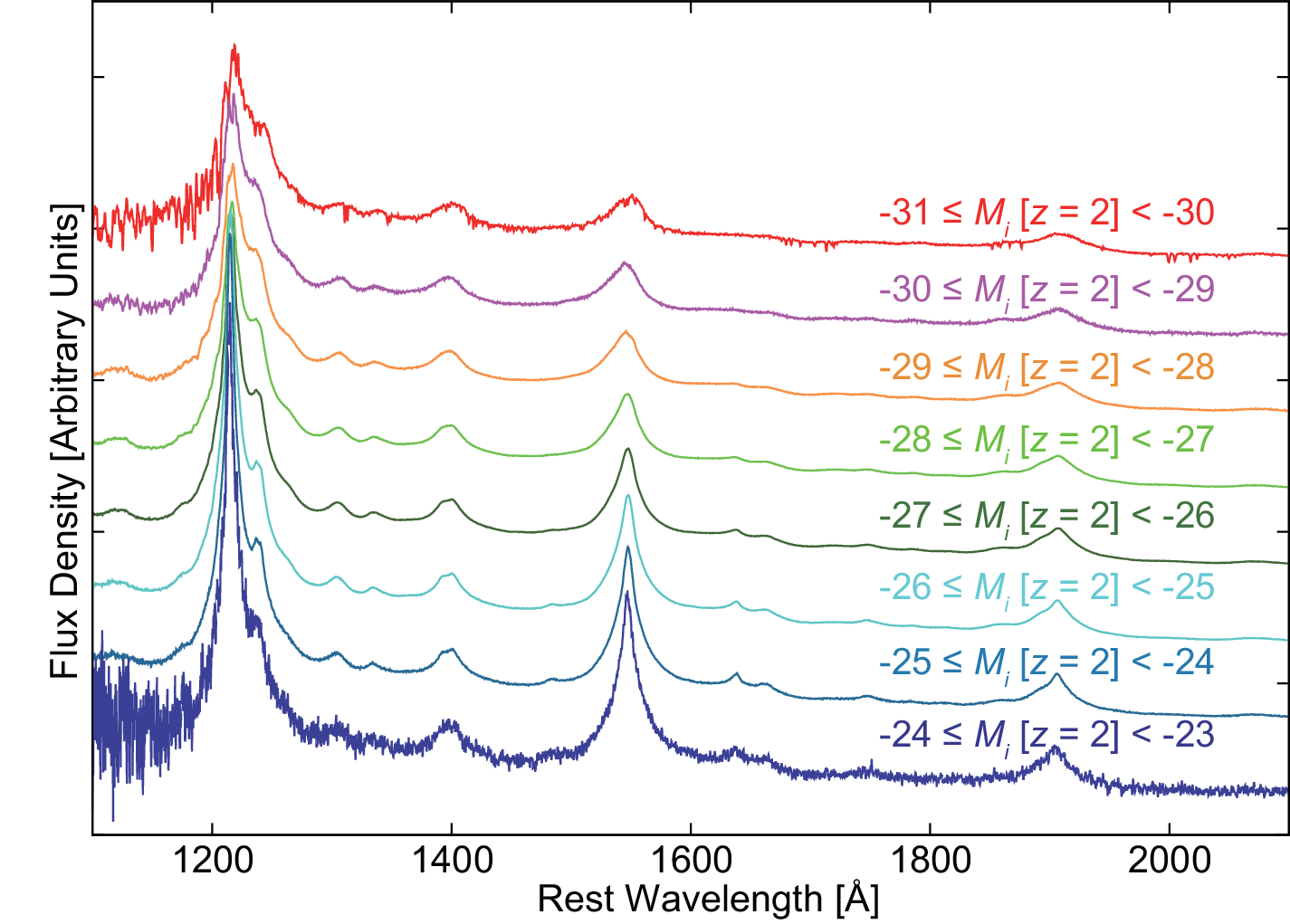}
\caption{
   The composite spectra of BOSS quasars for each absolute magnitude. The 
   difference of the line color denotes the difference of the luminosity. For clarifying 
   the difference of each composite spectrum in the figure, we shift each 
   composite spectrum toward the vertical direction.
   \label{fig1}}
\end{figure}

We then convert each spectrum from the observed frame to the rest frame, and then
normalize the each converted spectrum by averaging the flux at
1445$\ \rm{\AA}$ $<$ $\lambda_{\rm rest}$ $<$ 1485$\ \rm{\AA}$, which contains few 
emission-lines contributions \citep[see, e.g.,][]{van01, nag06}. Then 
we make the composite spectra for each absolute magnitude bin by using the IRAF 
\verb|scombine| task in the \verb|noao.|\verb|onedspec| package. The number of
quasars used to make the composite spectrum for each absolute magnitude is given
in Table~\ref{tbl-1}. In this stacking process, we adopt the $3\sigma$ clipping for 
removing the outlier pixels and then calculate the average flux at each wavelength in 
the logarithmic scale. Figure~\ref{fig1} shows the obtained composite spectra for 
each absolute magnitude in the rest frame. 

\begin{figure}
\epsscale{1.1}
\plotone{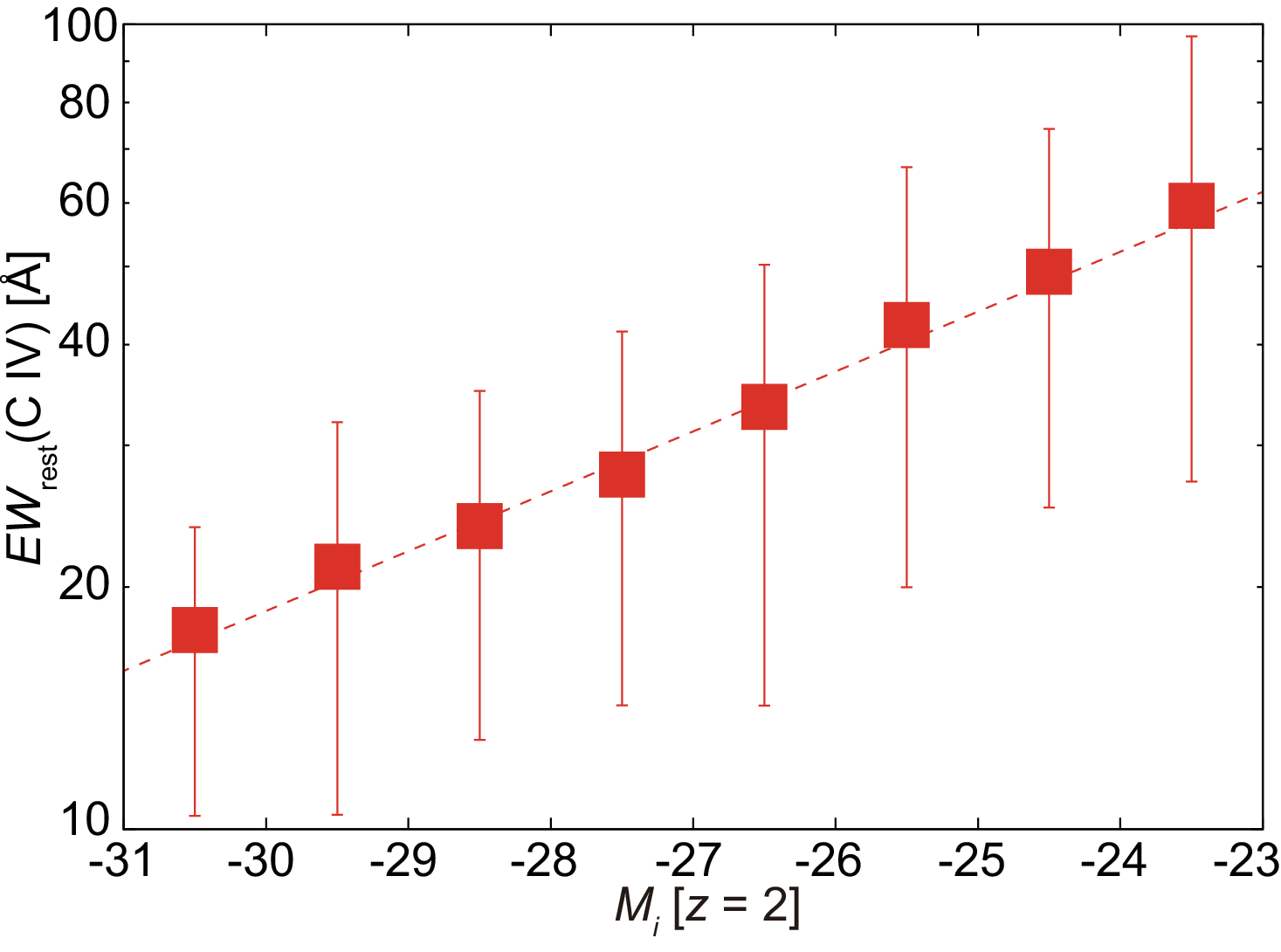}
\caption{
   The correlation between $EW_{\rm{rest}}$(C~{\sc iv}) measured from composite 
   spectra of BOSS quasars and their absolute magnitude. The 1$\sigma$ error bars 
   show the standard deviation of $EW_{\rm{rest}}$(C~{\sc iv}) of quasars in each 
   magnitude bin.  The dashed line shows the fit to the $EW_{\rm{rest}}$(C~{\sc iv}) 
   with the linear function.\label{fig2}}
\end{figure}

Figure~\ref{fig1} clearly shows a systematic trend that lower-luminosity quasars tend
to show emission lines with a larger $EW$. This is consistent with previous works of
the stacking analysis of high-redshift quasars \citep[e.g.,][]{nag06, mat11}. 
To investigate this trend quantitatively, we measure $EW$(C~{\sc iv}) 
for each composite spectrum by adopting the method of \citet{van01}. For 
subtracting the continuum flux, we define the local continuum around the C~{\sc iv} 
emission. This local continuum is defined by the median flux densities in the two 
wavelength regions, 1460$\ \rm{\AA}$ $<$ $\lambda_{\rm rest}$ $<$ 1475$\ \rm{\AA}$ 
and 1600$\ \rm{\AA}$ $<$ $\lambda_{\rm rest}$ $<$ 1630$\ \rm{\AA}$. Then we 
measure the flux of the C~{\sc iv} emission and the underlying continuum emission, 
that is converted to $EW_{\rm rest}$(C~{\sc iv}). We also derive the standard 
deviation of $EW_{\rm rest}$(C~{\sc iv}) for each absolute magnitude bin, by 
calculating the distribution of $EW_{\rm rest}$(C~{\sc iv}) in the individual (i.e., 
pre-stacking) spectrum of the BOSS quasars. 
The measurement results are given in 
Table~\ref{tbl-1}. We show the measured $EW_{\rm rest}$(C~{\sc iv}) as a function 
of the absolute magnitude of quasars in Figure~\ref{fig2}. A clear correlation between 
the $EW_{\rm rest}$(C~{\sc iv}) and the absolute magnitude is seen, that is 
consistent with the Baldwin effect.  We apply the weighted least-squares fit to the 
$EW_{\rm rest}$(C~{\sc iv}), adopting the following linear function: 
\begin{equation}
	{\rm log} [EW_{\rm rest}{\rm (C~IV)}] = {\rm A} \times M_i[z = 2] + {\rm B},
	\label{equ1}
\end{equation}
where A and B are fitting parameters. The fitting results tell us
${\rm A} = 0.074^{+0.003}_{-0.002}$ and ${\rm B} = 3.503^{+0.067}_{-0.096}$.
The fitting result is shown with the dashed line in Figure~\ref{fig2}.

\begin{deluxetable}{cccr}
\tabletypesize{\scriptsize}
\tablecaption{
   The measured $EW_{\rm{rest}}$({C~{\sc iv}}) of composite spectra of BOSS quasars 
   \label{tbl-1}}
\tablewidth{0pt}
\tablehead{
\colhead{$M_i[z = 2] $} & \colhead{$EW_{\rm{rest}}$({C~{\sc iv}}) } & 
\colhead{$\sigma$ ($EW_{\rm{rest}}$({C~{\sc iv}}))} & \colhead{$N$\tablenotemark{a}}\\
\colhead{$[\rm{mag}] $} & \colhead{[$\rm{\AA}$] } & 
\colhead{[$\rm{\AA}$]} & \colhead{}
}
\startdata
 $-31 \le M_i < -30$ & 17.7 & 6.67 & 5 \\
 $-30 \le M_i < -29$ & 21.2 & 10.8 & 84 \\
 $-29 \le M_i < -28$ & 23.8 & 11.1 & 1,047\\
 $-28 \le M_i < -27$ & 27.6 & 13.6 & 5,479\\
 $-27 \le M_i < -26$ & 33.5 & 18.0 & 14,424\\
 $-26 \le M_i < -25$ & 42.3 & 23.2 & 17,719\\
 $-25 \le M_i < -24$ & 49.3 & 24.5 & 5,131\\
 $-24 \le M_i < -23$ & 59.5 & 34.8 & 64
\enddata
\tablenotetext{a}{Number of quasars in each magnitude bin.}
\end{deluxetable}

\subsection{Colors of Quasars} \label{subsec:2.3}

To examine how the color of quasars is affected by the Baldwin effect inferred in
Section~\ref{subsec:2.2}, we make simple model spectra for each quasar luminosity
following previous works \citep[e.g.,][]{fan99, hun04, ric06, sia08, ike11, ike12}.  
Here we assume that the continuum spectral energy distribution (SED) of quasars 
does not depend on the redshift \citep[see, e.g.,][]{kuh01, tel02, yip04, jia06}.
Because \citet{tel02} show that the continuum SED of radio-quiet quasars is 
independent of the luminosity, we also assume that the continuum SED  of quasars 
does not depend on the luminosity. We adopt the double power-law continuum 
($f_\nu \propto \nu^{-\alpha_{\nu}}$), showing a spectral break at $\lambda_{\rm rest}$ 
= 1100\ \AA. The spectral slope at the shorter wavelength side is $\alpha_{\nu} = 1.76$  
\citep{tel02} while that at longer wavelength side is $\alpha_{\nu} = 0.46$ \citep{van01}.
On this continuum emission, strong emission lines are added by adopting the
measured $EW_{\rm{rest}}$(C~{\sc iv}) given in Table~\ref{tbl-2} and the typical
emission-line flux ratios given in Table~2 of \citet{van01}. Here we include only
emission lines whose flux is larger than $0.5$\% of the Ly$\alpha$ flux. We also
include the Balmer continuum and the Fe~{\sc ii} multiplet emission features by 
adopting the template give by \citet{kaw96}. The relative strength of the emission
lines to the continuum emission is determined by $EW_{\rm rest}$(C~{\sc iv})
measured in Section~\ref{subsec:2.2}, depending on the quasar absolute
magnitude (Table~\ref{tbl-1}).
The effects of the intergalactic absorption by the neutral hydrogen are corrected by 
adopting some extinction models. As for the extinction model, a widely used one
is the model of \citet{mad95}. However, \citet{ino14} recently pointed out that the 
model of \citet{mad95} overestimates the IGM attenuation for $z \sim 4-5$.
Therefore we also use the extinction model proposed recently by \citet{ino14} in
addition to the model of \citet{mad95} for investigating how the adopted model
for the intergalactic absorption affects the quasar colors.

\begin{figure}
\epsscale{1.1}
\plotone{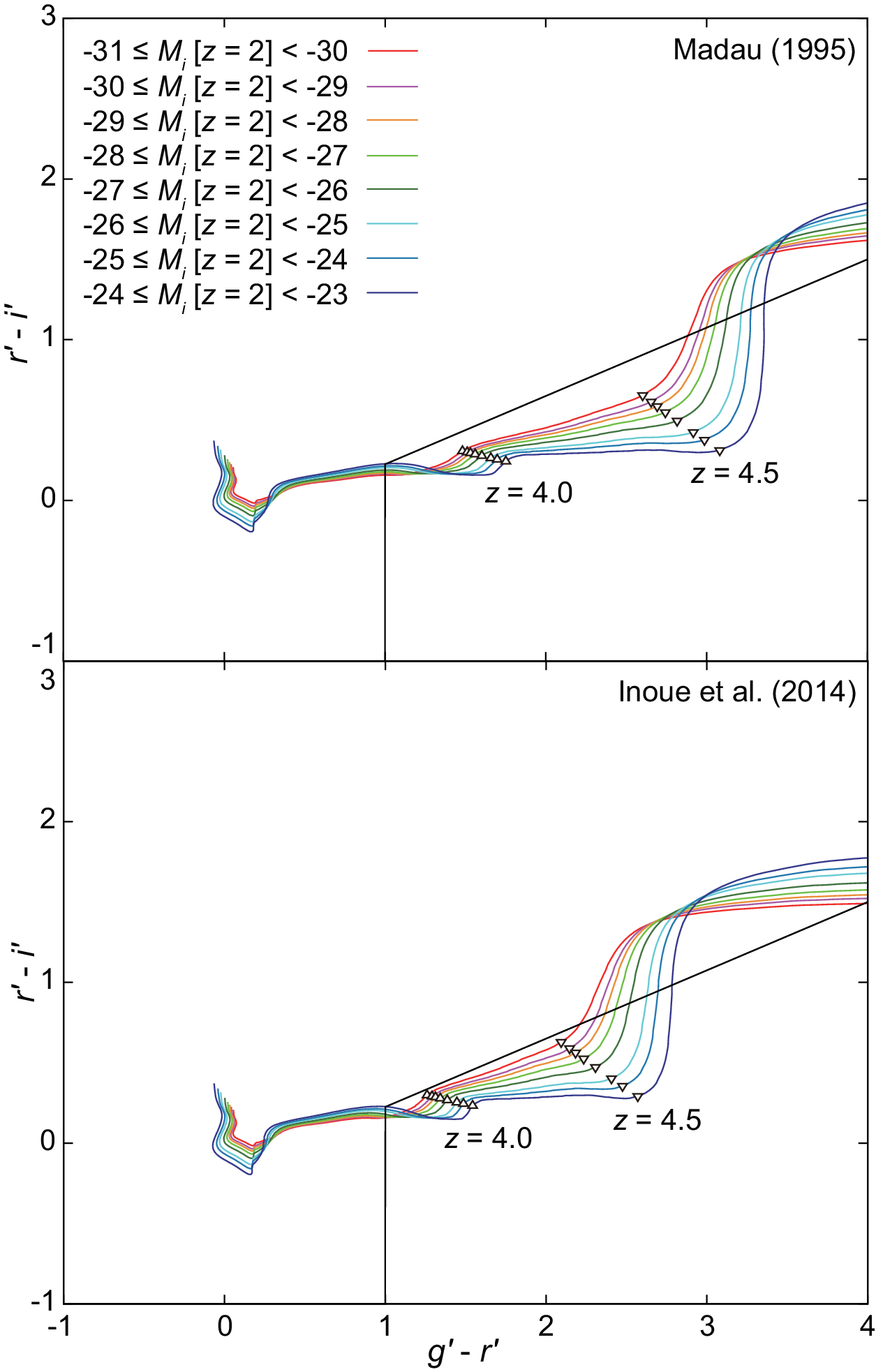}
\caption{
   Two-color diagram of $r' - i'$ versus $g' - r'$. The color tracks of the model 
   quasar adopting the IGM absorption model of \citet{mad95} are shown in the 
   upper panel, while those adopting the model of \citet{ino14} are shown in the 
   lower panel. Different colors denote the different absolute magnitude for
   each model of the quasar spectrum, as explained in the upper panel. The 
   photometric criteria adopted for selecting photometric quasar candidates at 
   $z \sim 4$ (\citealt{ike11}) are shown by the black solid lines.
\label{fig3}
}
\end{figure}

\begin{figure}
\epsscale{1.1}
\plotone{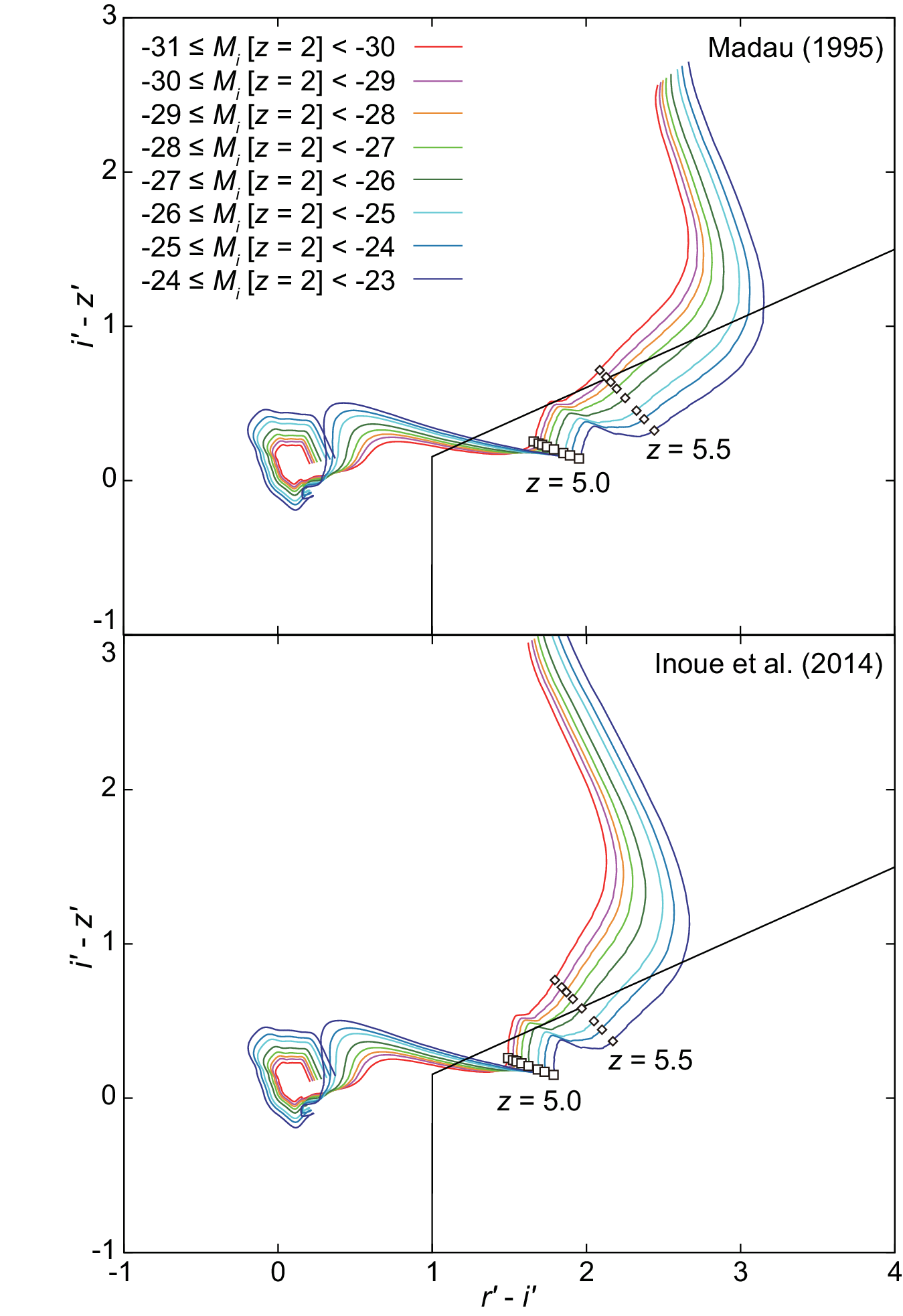}
\caption{
   Two-color diagram of $i' - z'$ versus $r' - i'$. The color tracks of the model 
   quasar adopting the IGM absorption model of \citet{mad95} are shown in the 
   upper panel, while those adopting the model of \citet{ino14} are shown in the 
   lower panel. Colored lines are the same as in Figure~\ref{fig3}. The 
   photometric criteria adopted for selecting photometric quasar candidates at 
   $z \sim 5$ (\citealt{ike12}) are shown by the black solid lines.
\label{fig4}
}
\end{figure}

\begin{figure}
\epsscale{1.1}
\plotone{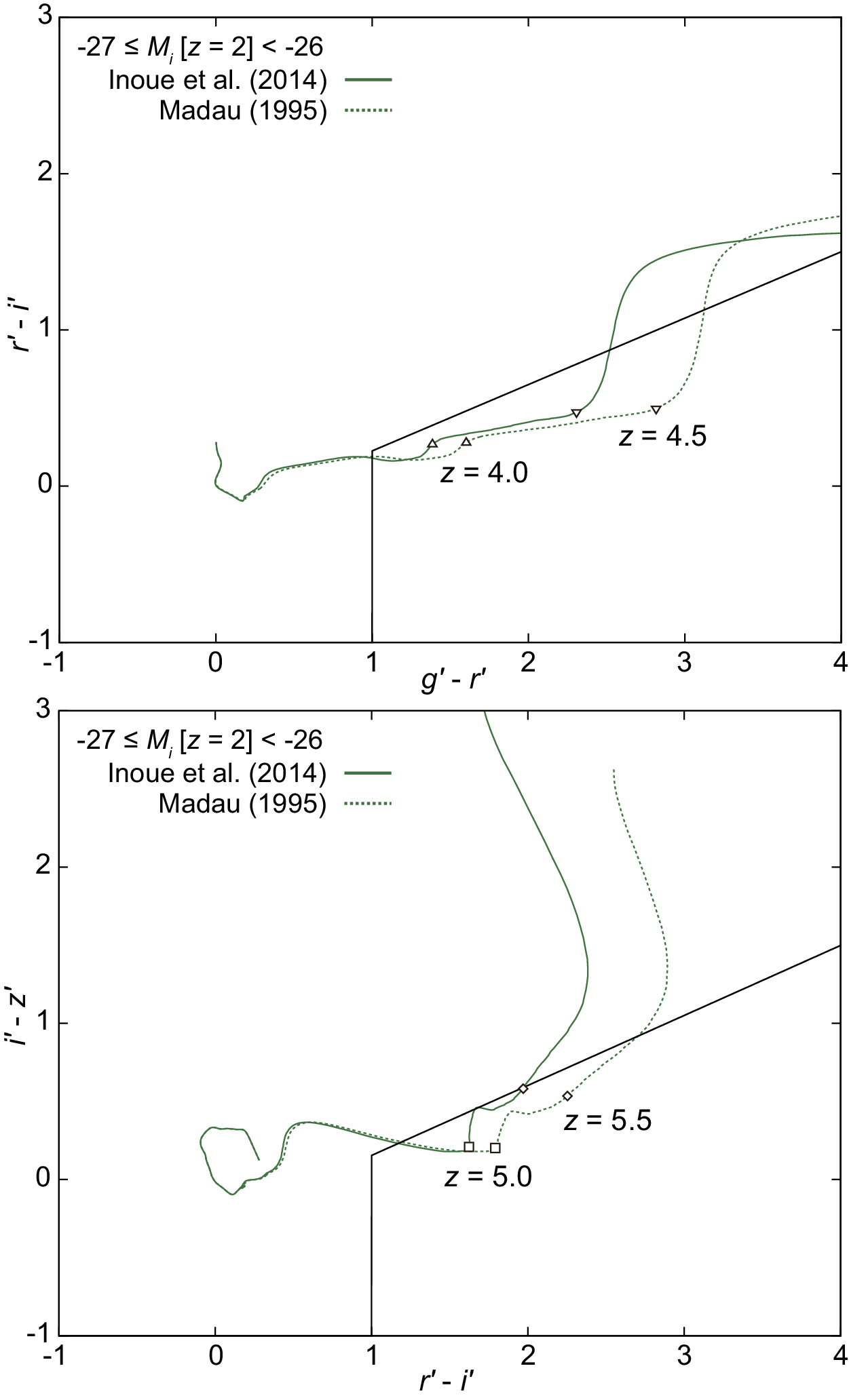}
\caption{
   Two-color diagrams of $r' - i'$ versus $g' - r'$ (upper panel) and $i' - z'$ versus 
   $r' - i'$ (lower panel). The solid and dotted lines denote the color tracks of the 
   model quasar with $-27 \le M_i[z=2] < -26$, adopting the IGM absorption
   model of \citet{ino14} and that of \citet{mad95} respectively. The 
   photometric criteria adopted for selecting photometric quasar candidates at 
   $z \sim 4$ (\citealt{ike11}) and $z \sim 5$ (\citealt{ike12}) are shown by the 
   black solid lines in the upper and lower panels, respectively.
\label{fig5}
}
\end{figure}

Based on the prepared spectral model, we calculated the $g' - r'$, $r' - i'$, 
and $i' - z'$ colors of model quasars with different absolute magnitudes (showing
different $EW$ of emission lines) at redshifts from 2 to 6. Here we use the filter 
response function of Suprime-Cam \citep{miy02} boarded on the Subaru telescope, 
because we will compare our results with previous observations using
Suprime-Cam in the later sections. The simulated colors of the model quasars for 
each absolute magnitude are shown in the $r' - i'$ versus $g' - r'$ diagram 
(Figure~\ref{fig3}) and the $i' - z'$ versus $r' - i'$ diagram (Figure~\ref{fig4}). To
examine the effect of the adopted IGM absorption model, we show the colors of 
the model quasar spectra adopting the model of \citet{mad95} and those adopting
the model of \citet{ino14} separately in those figures. For investigating the effect 
of adopting different IGM absorption models more evidently, the color tracks for 
quasars with a specific absolute magnitude ($-27 \le M_i[z = 2] < -26$) but adopting
two different IGM absorption models are shown in the 2 two-color diagrams
(Figure~\ref{fig5}). The quasar selection criteria in these two-color diagrams adopted 
by \citet{ike11} and \citet{ike12} are also shown in Figures~\ref{fig3}--\ref{fig5}. 
Specifically, the criteria for selecting quasars at $z \sim 4$ are
\begin{equation}
	r' - i' < 0.42 (g' - r') - 0.22,
	\label{equ2}
\end{equation}
\begin{equation}
	g' - r' > 1.0,
	\label{equ3}
\end{equation}
and the criteria at $z \sim 5$ are
\begin{equation}
	i' - z' < 0.45 (r' - i') - 0.24,
	\label{equ4}
\end{equation}
\begin{equation}
	r' - i' > 1.0. 
	\label{equ5}
\end{equation}

Figures \ref{fig3} and \ref{fig4} clearly show that quasars with a lower luminosity 
satisfy more easily the selection criteria than quasars with a higher luminosity, 
both at $z \sim 4$ and $z \sim 5$. This is mainly because the redshifted 
Ly$\alpha$ emission (whose $EW$ is larger in lower-luminosity quasars) locates 
at the wavelength within the $r'$-band coverage at $z \sim 4$ and within the 
$i'$-band coverage at $z \sim 5$. This result suggests that previous optical 
photometric surveys for high-redshift low-luminosity quasars overestimated the
quasar number density if a spectral template made from bright quasars (such as
the composite spectrum given by \citealt{van01}) is used, due to the under 
estimation of the survey completeness. As for the effect of the adopted IGM 
absorption model, it is clearly shown in Figures~\ref{fig3}--\ref{fig5} (especially in 
Figure ~\ref{fig5}) that the separation between the color tracks and the selection 
boundary in the two-color diagrams is smaller when we adopt the model of 
\citet{ino14} than the case when we adopt the model of \citet{mad95}. This 
suggests that previous high-redshift quasar surveys adopting the IGM absorption 
model by \citet{mad95} underestimate the quasar number density due to the over 
estimation of the survey completeness.
Accordingly, it is interesting to assess how the previous quasar surveys
actually overestimated or underestimated the number density of high-redshift 
quasars quantitatively. Therefore, in the next section, we investigate the effects 
of the luminosity dependence of the quasar spectral features and also the 
adopted IGM absorption model on the estimates of the survey completeness.

\section{RESULTS}\label{sec:3}

\subsection{The COSMOS Data} \label{subsec:3.1}

Here we study how the luminosity dependence of the quasar spectrum and the
selection of the IGM absorption model affect the estimate of the survey 
completeness. For this purpose, we specifically focus on a previous survey for
low-luminosity quasars at $z\sim4$ and $z\sim5$ \citep{ike11, ike12}. This survey
was conducted by using the dataset obtained in the Cosmic Evolution Survey
(COSMOS) field. The COSMOS is a treasury program of the Hubble Space 
Telescope (HST) \citep{sco07}. The COSMOS field covers an area of $1^{\circ}.4 
\times 1^{\circ}.4$ which corresponds to $\sim$2 deg$^2$, 
centered at R.A. (J2000) = 10:00:28.6 and Dec. (J2000) = +02:12:21.0.  

For examining the completeness of high-redshift quasar surveys, we use the 
optical broad-band photometric data in the COSMOS field obtained with 
Suprime-Cam boarded on the Subaru telescope. In this study we focus on quasars
at $z\sim4$ and $z\sim5$, that are selected through $g'$-band dropout and
$r'$-band dropout technique. 
Therefore we use the Suprime-Cam images obtained with the $g', r', i'$, and 
$z'$-band filters \citep{tan07}. The $5\sigma$ depth in 3\arcsec aperture magnitudes 
are $g' = 26.5$, $r' = 26.6$, $i' = 26.1$, and $z' = 25.1$ \citep{cap07}.

\subsection{Completeness for Quasar Selection} \label{subsec:3.2}

The completeness in quasar surveys is generally defined as the number ratio of the 
photometrically selected quasars to the existing quasars, as functions of the
apparent magnitude and the redshift. We can derive this completeness by simulating 
the observation through a Monte Carlo approach; i.e., preparing artificially-prepared 
quasars based on the spectral modeling, putting the simulated quasar into the 
observational image with photometric errors, detecting them in the same manner as 
usual object detections, measuring the colors of the detected objects through the
forced photometry, applying the quasar selection criteria, and then obtain the 
completeness by calculating the number ratio of the selected model quasars to the 
prepared model quasars. In this work, we first estimate the completeness by 
adopting quasar spectral templates with different emission-line $EW$s (that 
corresponds to different absolute magnitudes as given in Table~\ref{tbl-1}) as a 
function of redshift, and then we derive the completeness as a function of the
apparent $i$-band magnitude and redshift.  
 
We generate the model quasar spectra in a similar way as described in 
Section~\ref{subsec:2.3}. Here we also take into account the intrinsic variation in the 
continuum slope and $EW$s of the emission lines. We assume Gaussian distributions 
of the power-law slope $\alpha_{\nu}$ ($f_{\nu} \propto \nu^{-\alpha_{\nu}}$) and 
C~{\sc iv} $EW$s. The mean of the slope is 1.76 at the bluer wavelength than the 
spectral break at 1100$\ \rm{\AA}$ in the rest frame, and 0.46 at the redder wavelength 
than the break (the same as those in Section~\ref{subsec:2.3}). The standard deviation 
of the slope is $0.30$ at all wavelengths \citep{fra96, hun04, tel02, ike11, ike12}.   
We then add emission-line features including the Balmer continuum and Fe~{\sc ii}
multiplet emission in the same way as described in Section~\ref{subsec:2.3}, taking 
account of the luminosity dependence of the emission-line $EW$ (Table~\ref{tbl-1}).
Here we also take the intrinsic variation in the emission-line strength into account,
whose standard deviation depends on the quasar luminosity as given in 
Table~\ref{tbl-1}. In this way we create 1,000 quasar spectra for each composite 
spectrum at each $\Delta z = 0.01$ in the redshift range of $3.4 \le z \le 6.0$. 
The effects of the intergalactic absorption 
are corrected by adopting the extinction model of \citet{ino14}. 

By utilizing the realized spectrum of each model quasar, we calculate $g'$, $r'$, and 
$z'$-band apparent magnitudes of the model quasar based on $i'$-band apparent 
magnitude ($22 \le i' \le 24$, $\Delta i' = 0.5$). And we derive the colors of each 
model quasars ($g' - r'$, $r' - i',$ and $i' - z'$) in the observed frame. To estimate the 
photometric completeness, we put the 1000 model quasars for each parameter set
(the apparent magnitude with $\Delta i' = 0.5$, redshift with $\Delta z = 0.01$, and the
absolute magnitude of the adopted template with $\Delta M_i [z = 2] = 1.0$) into the 
COSMOS FITS images taken with Suprime-Cam as point sources. For this process,
we use the IRAF \verb|mkobjects| task in the \verb|artdata| package. Then we try to
detect them in the $i'$-band image with SExtractor \citep{bear96}, and measure their
colors with the double-image mode. During the process of the source photometry,
we correct the photometric offset of each image by refering to \citet{cap07}.

\begin{figure*}
\epsscale{1.1}
\plottwo{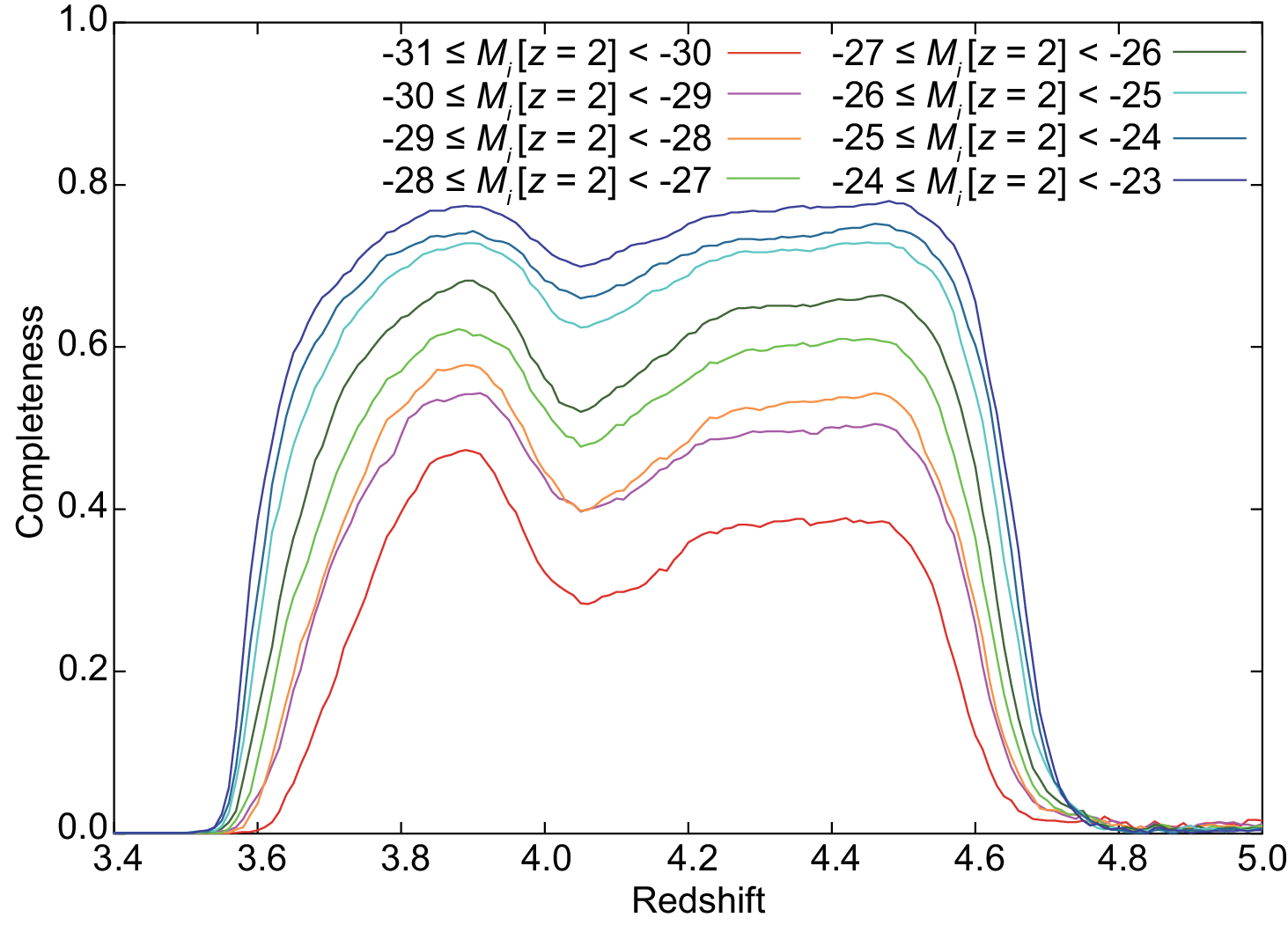}{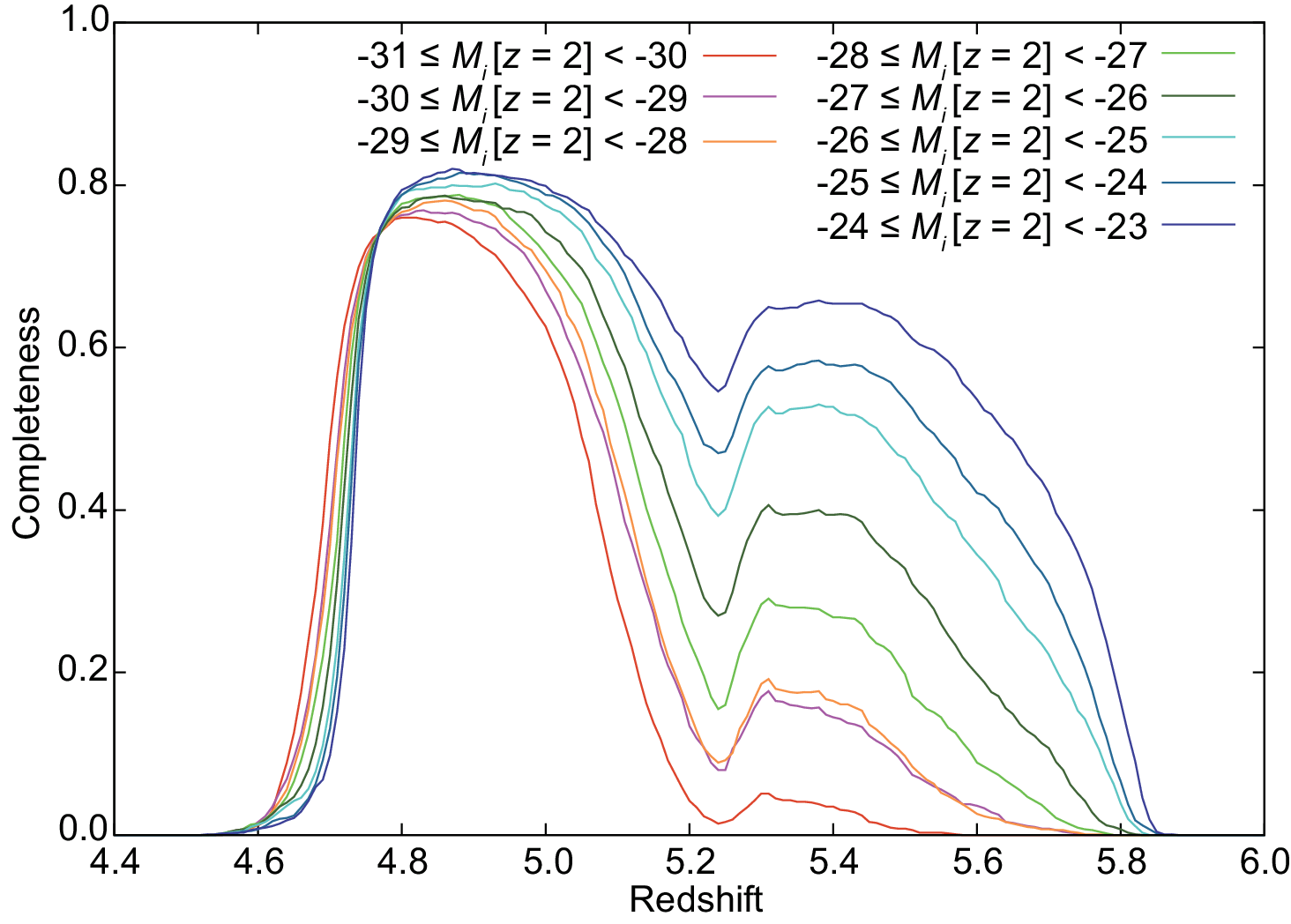}
\caption{
     The photometric completeness for COSMOS quasars with $i' = 23$ but having
     different emission-line $EW$s (denoted with different colors) at $z \sim 4$ (left) 
     and at $z \sim 5$ (right). Note that the absolute magnitude given for each
     color code is just for characterizing the spectral template but is not related 
     to the signal-to-noise ratio of simulated quasars.
     \label{fig6}}
\end{figure*}

The measured apparent magnitudes and colors of the simulated quasars are 
generally different from those before inserted into the Suprime-Cam images, due to 
the effects of photometric errors and neighboring foreground (or background) objects.
Accordingly, some model quasars are not selected as photometric candidates of 
quasars with the adopted criteria (see Section~\ref{subsec:2.3}), that causes the
degrade of the survey completeness. The survey completeness is derived by 
calculating the fraction of model quasars that are selected as photometric candidates 
in the above process. Figure~\ref{fig6} shows the derived photometric completeness 
for quasars with $i' = 23$ for $z \sim 4 - 5$, by adopting different spectral templates 
with different emission-line $EW$s characterized by the $i$-band absolute magnitude, 
$M_i [z = 2]$. Note that the absolute magnitude defining the spectral template is 
inconsistent to the combination of the redshift and the assumed apparent magnitude 
($i' = 23$); this is because we intend to investigate how the completeness is sensitive 
to the adopted quasar template (emission-line $EW$s) first.
The derived completeness highly depends on the adopted spectral template in the
sense that the completeness is lower when the spectral template made of 
higher-luminosity quasars (i.e., with smaller emission-line $EW$) is adopted, even
though the same $i'$-band apparent magnitude ($i' = 23$ for Figure~\ref{fig6}) is 
assumed. This is consistent to Figures~\ref{fig3} and \ref{fig4}, where it is inferred 
that lower-luminosity quasars satisfy the photometric selection criteria more easily.
Note that a small dip in the completeness curve is seen at $z \sim 3.9 - 4.2$ and
$z \sim 5.2 - 5.3$ in Figure~\ref{fig6}, that is due to the flux contribution of the 
C~{\sc iv} emission that makes the $r' - i'$ and $i' - z'$ colors redder at each redshift 
range respectively.

We then derive the completeness as a function of the $i$-band apparent magnitude
and redshift, by utilizing the results shown in Figure~\ref{fig6}. Since the combination
of the apparent magnitude and redshift tells us which spectral template 
(characterized by the $i$-band absolute magnitude) should be used for the estimate 
of the completeness in terms of the absolute magnitude.
For this purpose, we convert the $i'$-band apparent magnitude to the absolute AB 
magnitude at 1450$\ \rm{\AA}$, $M_{1450}$, \citep[e.g.,][]{ric06, cro09, gli10} 
through the following relation: 
\begin{eqnarray}
   M_{1450} = 
   &m_{i'}&  + \ 5 - 5{\rm log}d_L(z) + 2.5(1 - \alpha_{\nu}){\log}(1 + z) \nonumber \\
   &+& 2.5\alpha_{\nu}{\rm log}(\frac{\lambda_{i'}}{1450\ \rm{\AA}}),
   \label{equ6}
\end{eqnarray}
where $d_L(z)$, $\alpha_{\nu}$, and $\lambda_{i'}$ are the luminosity distance, 
spectral index of the quasar continuum ($f_{\nu} \propto \nu^{-\alpha_{\nu}}$), and 
the effective wavelength of the Suprime-Cam $i'$-band ($\lambda_{i'} = 7684\ \rm{\AA}$), respectively. We assume $\alpha_{\nu} = 0.46$ in Equation~(\ref{equ6}).  
Also we convert $M_i [z = 2]$ to $M_{1450}$, \citep{ric06, ros13} through the 
following relation:
\begin{eqnarray}
   M_i[z = 2] = & M_{1450} & + 2.5 \alpha_{\nu} \rm {log} 
   (\frac{1450\ \AA}{\lambda_{\it{i}^{\ast}}}) \nonumber \\
   & - & 2.5 (1 - \alpha_{\nu}) \rm{log}(1+2),
   \label{equ7}
\end{eqnarray}
where $\alpha_{\nu}$ and $\lambda_{i^{\ast}}$ are the spectral index of the quasar 
continuum (0.46) and the effective wavelength of the SDSS $i^{\ast}$-band 
($\lambda_{i^{\ast}} = 7471\ \rm{\AA}$), respectively.
Then we derived the completeness for each combination of the redshift and
the apparent $i'$-band magnitude using an appropriate spectral template 
characterized by $M_i [z = 2]$.

\begin{figure*}
\epsscale{1.1}
\plottwo{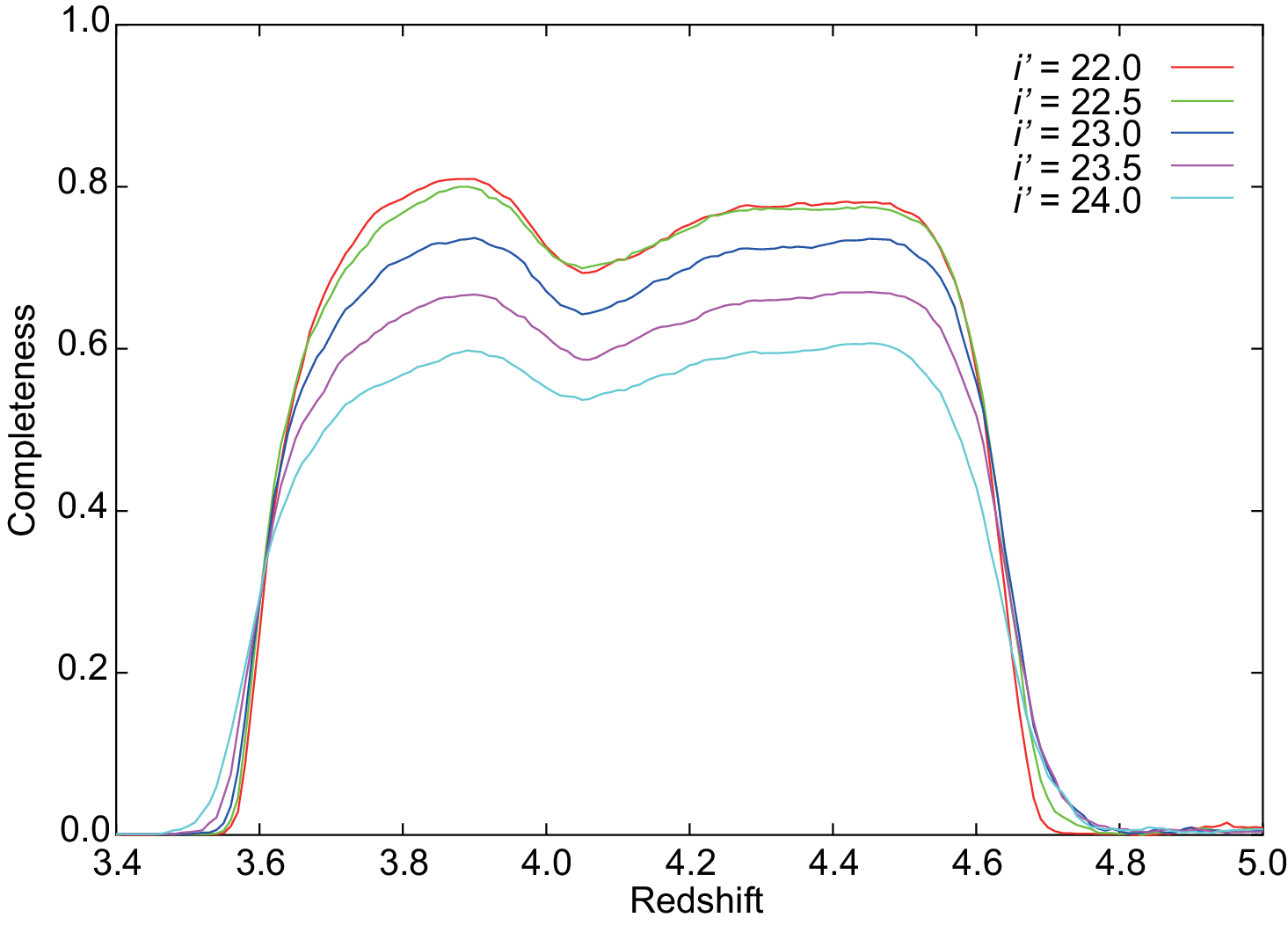}{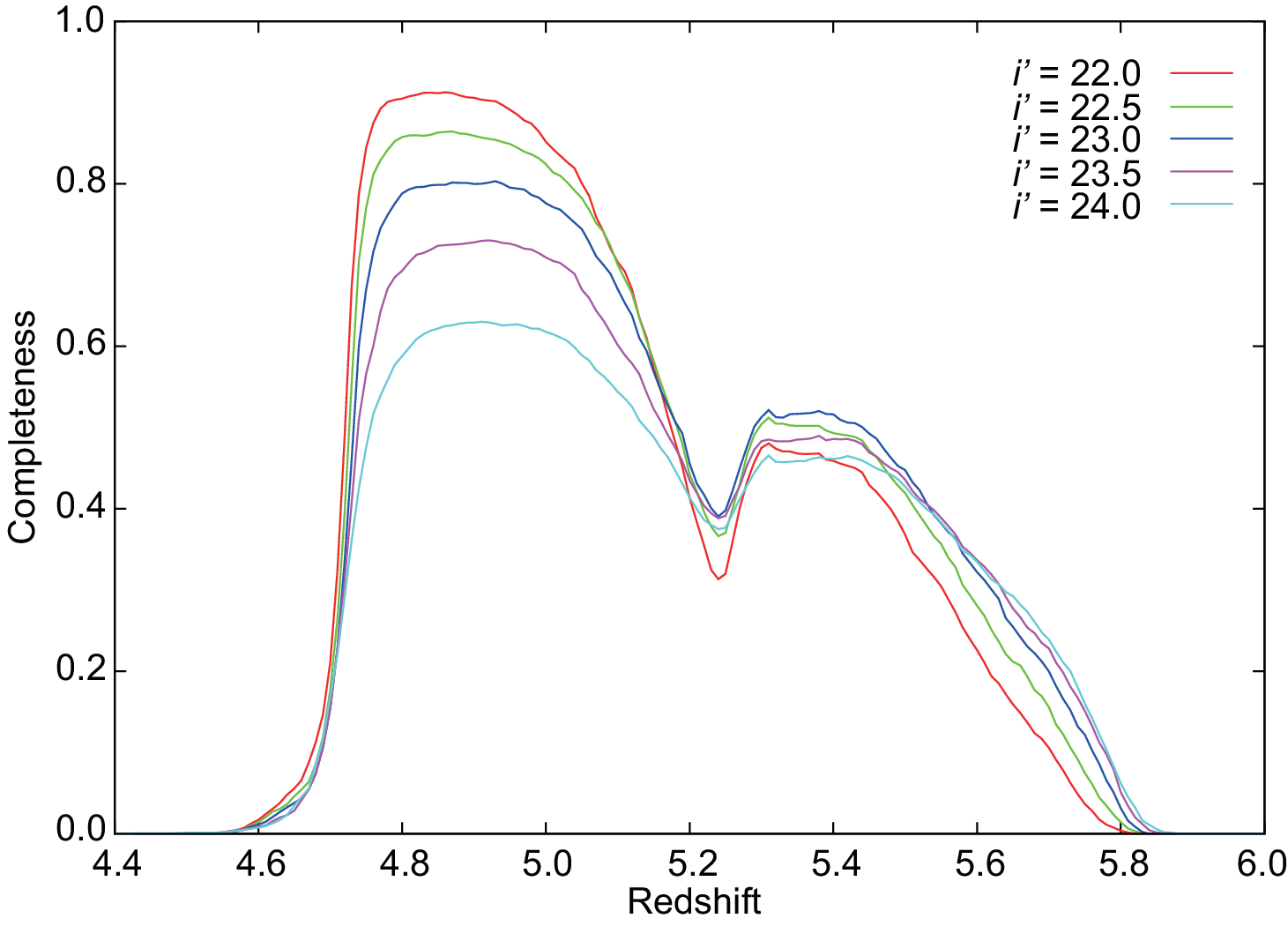}
\caption{
     The photometric completeness for COSMOS quasars at $z \sim 4$ (left) and 
     at $z \sim 5$ (right). Red, green, blue, purple, and cyan lines denote the 
     estimated completeness for quasars with $i' = 22.0, 22.5, 23.0, 23.5, \rm {and} 
     \ 24.0$, respectively. \label{fig7}}
\end{figure*}

Figure~\ref{fig7} shows the photometric completeness as a function of the redshift 
and the apparent $i'$-band magnitude. Although the derived completeness is 
generally higher for brighter $i'$-band magnitudes as expected, an invert tendency 
is seen at $z \gtrsim 5.2$. This is due to the luminosity dependence of quasar 
spectrum, that makes the completeness higher for lower-luminosity quasars (see 
Section~\ref{subsec:2.3}). The estimated completeness at $z \sim 4$ is generally 
higher than that estimated by \citet{ike11}, which is also due to the luminosity 
dependence of quasar spectra since \citet{ike11} used a composite spectrum of 
bright SDSS-selected quasars \citep{van01} for the completeness estimate.
However at $z \sim 5$, the estimated completeness in this work is lower than
that estimated by \citet{ike12}. This is because the new extinction model
\citep{ino14} makes the completeness lower especially at $z \sim 5$, whose 
effect gives stronger impact on the completeness estimate than the effect of 
the luminosity dependence of the quasar spectrum.

\section{DISCUSSION}\label{sec:4}
 
\subsection{The QLF at $z \sim 4$}\label{subsec:4.1}
 
Based on the revised completeness for the previous quasar survey through the 
optical color selection at the COSMOS field \citep{ike11, ike12}, we revisit the
QLF at $z \sim 4 - 5$. First we compute the effective comoving volume of the 
COSMOS quasar survey as
\begin{equation}
     V_{\rm {eff}}(m_{i'}) = 
     d\Omega \int_{z = 0}^{z = \infty} C(m_{i'}, z) \frac{dV}{dz}dz,
     \label{equ8}
\end{equation}
where $d\Omega$ (1.64 deg$^2$) is the solid angle of the survey and $C(m_{i'}, z)$ is the 
photometric completeness studied in Section~\ref{subsec:3.2}. In \citet{ike11}, 
eight spectroscopically-confirmed quasars in the COSMOS field are used to derive 
the QLF at $z \sim 4$.  
Taking into account of the possibility that some of quasar photometric candidates 
without spectra could also be quasars (that is estimated based on the success rate 
of the spectroscopic run), we statistically adopt the number of quasars in the 
COSMOS survey at $z \sim 4$ as follows; 4.67 for $-24.09 < M_{1450} < -23.09$ 
and 6.06 for $-23.09 < M_{1450} < -22.09$, respectively \citep{ike11}.  
By using these corrected numbers of quasars and the effective comoving volume, 
we calculate the space density of quasar and their standard deviation as 
 \begin{equation}
     \Phi(m_{i'}, z) = \sum_{\rm{j}}\frac{1}{V_{\rm{eff}}^{\rm{j}}\Delta m_{i'}} = \frac{N_{\rm{cor}}}{V_{\rm{eff}}},
       \label{equ9}
\end{equation}
\begin{equation}
    \sigma(\Phi) = \sum_{\rm{j}}[(\frac{1}{V_{\rm{eff}}^{\rm{j}}\Delta m_{i'}})^2]^{1/2} = [N_{\rm{cor}}(\frac{1}{V_{\rm{eff}}})^2]^{1/2},
     \label{equ10}
\end{equation} 
where $\Delta m_{i'} = 1$, $N_{\rm{cor}}$ is the corrected number of quasars.
The calculated space density of quasar and their standard deviation are
$(3.49 \pm 1.62) \times 10^{-7}$ Mpc$^{-3}$ mag$^{-1}$ for 
$-24.09 < M_{1450} < -23.09$ and 
$(5.24 \pm 2.13) \times 10^{-7}$ Mpc$^{-3}$ mag$^{-1}$ for 
$-23.09 < M_{1450} < -22.09$, respectively.  

\begin{figure}
\epsscale{1.1}
\plotone{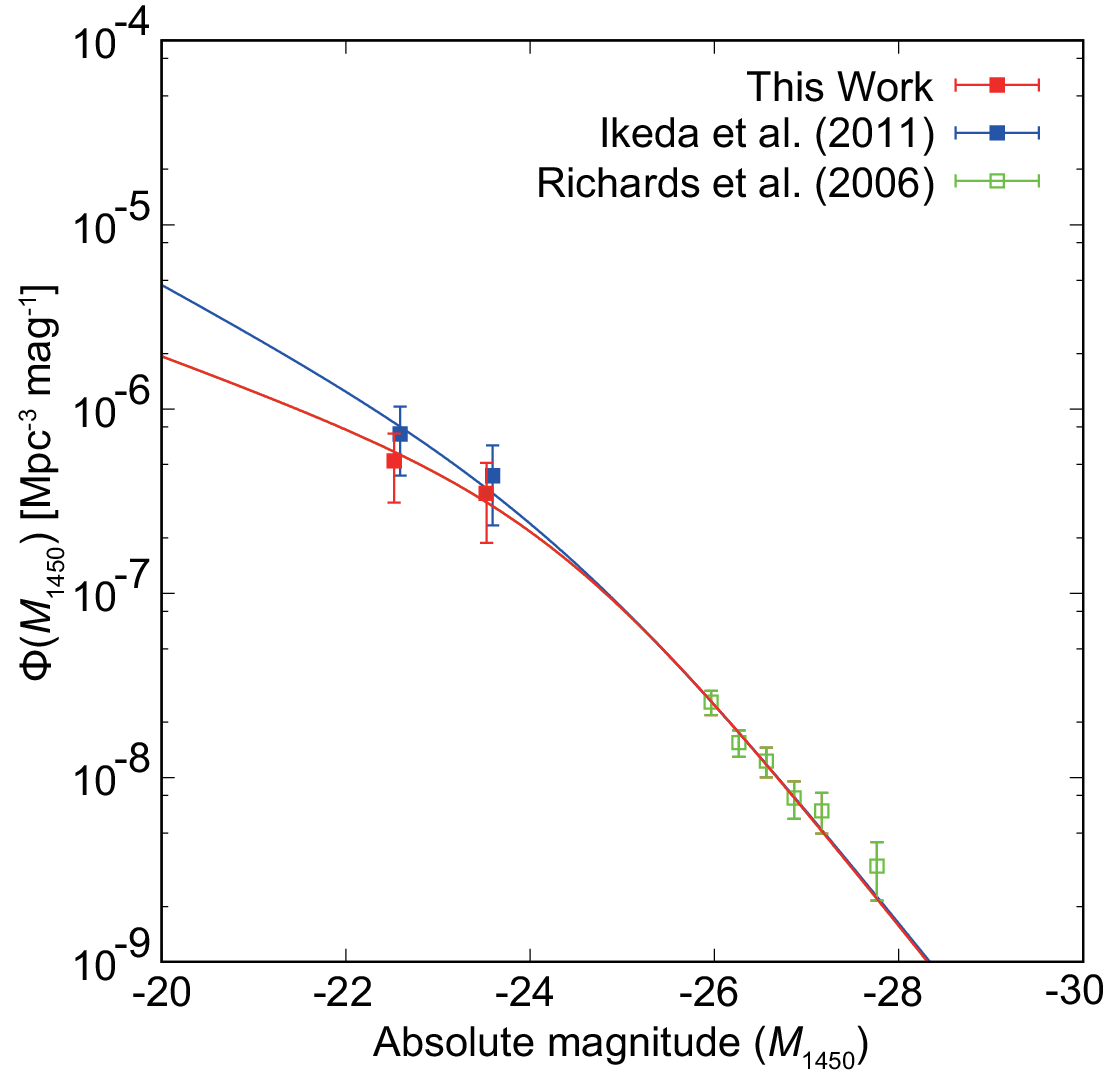}
\caption{
     The QLF at $z \sim 4$ derived from the COSMOS data. 
     The red and blue squares are the quasar number density calculated in this work
     and in \citet{ike11}, respectively. The red squares are slightly shifted to the left 
     direction to avoid the overlap with the blue squares. Green squares are the SDSS 
     results \citep{ric06}. The red line shows the fit to our results and the SDSS data,
     while the blue line shows the fit to the result of \citet{ike11} and of the SDSS. 
     \label{fig8}}
\end{figure}

The corrected number and calculated space density of $z \sim 4$ quasars in
COSMOS are showed in Table~\ref{tbl-2}, and the obtained QLF is plotted in 
Figure~\ref{fig8} with the results of \citet{ike11} and SDSS \citep{ric06}.
Although the redshift range of SDSS data \citep[$4.0 \le z \le 4.5$;][]{ric06} is slightly 
different from our study ($3.7 \le z \le 4.7$), we apply the weighted least-squares fit 
to the space density of quasars at $z \sim 4$ inferred by both our study and SDSS.
Here the following double power-law function is adopted:
\begin{equation}
     \tiny{\Phi(M_{1450}, z) = \frac{\Phi(M^*_{1450})}
     {10^{0.4(\alpha + 1)(M_{1450} - M^*_{1450}) }+ 10^{0.4(\beta + 1)(M_{1450} - M^*_{1450})}},}
     \label{equ11}
\end{equation}
where $\alpha$, $\beta$, $\Phi(M^{\ast}_{1450})$, and $M^{\ast}_{1450}$ are the 
bright-end slope, the faint-end slope, the normalization of the luminosity function, and 
the characteristic absolute magnitude, respectively \citep{boy00}. Among the four parameters, the 
bright-end slope is fixed to be $\alpha = -2.58$ based on the SDSS result at $z \sim 4$ 
\citep{ric06}. 
Since the characteristic absolute magnitude becomes too bright when the 
parameter is treated as a free parameter in the fit, the characteristic magnitude is fixed to be 
$M^{\ast}_{1450} = -24.4$ based on the result of the COSMOS QLF at $z \sim 4$ 
\citep{ike11}.
The best-fit values and 1$\sigma$ standard deviations of
$\Phi(M^{\ast}_{1450})$ and $\beta$ are $\Phi(M^{\ast}_{1450}) = (3.03 \pm 0.24) 
\times10^{-7}$ Mpc$^{-3}$ mag$^{-1}$ and $\beta = -1.46^{+0.23}_{-0.19}$, 
respectively. These best-fit results are showed in Table~\ref{tbl-3}, and the fitting 
results are shown in Figure~\ref{fig8}. Comparing with the results by \citet{ike11}, we 
find that our results taking account of the Baldwin effect and the IGM model show $\sim$24\% lower 
number density at $z \sim 4$. This is attributed by the revision of the completeness 
estimate and the adopted latest IGM absorption model from the previous work 
\citep{ike11}. Note that the revised number density of low-luminosity quasars at $z \sim 4$ is consistent with the number density inferred by recent X-ray observations \citep{mar16}.

\begin{deluxetable}{cccc}
\tabletypesize{\scriptsize}
\tabletypesize{\scriptsize}
\tablecaption{The quasar space density at $z \sim 4$ and their standard deviation\label{tbl-2}}
\tablewidth{0pt}
\tablehead{
\colhead{}& \colhead{} & \colhead{This Work} & \colhead{\citet{ike11}}\\ 
\colhead{$M_{1450}$} & \colhead{$N_{\rm {cor}}$\tablenotemark{a}} & 
\colhead{$\Phi$} & \colhead{$\Phi$}\\
 \colhead{[mag]} & &  \colhead{[$10^{-7}$ Mpc$^{-3}$ mag$^{-1}$]} &  \colhead{[$10^{-7}$ Mpc$^{-3}$ mag$^{-1}$]}
}
\startdata
-23.59 & 4.67 & 3.49 $\pm{1.62}$ & 4.35 $\pm{2.01}$\\
-22.59 & 6.06 & 5.24 $\pm{2.13}$ & 7.32 $\pm{2.97}$
\enddata
\tablenotetext{a}{$N_{\rm {cor}}$ is the corrected number of the quasars \citep{ike11}.}
\end{deluxetable}

\begin{deluxetable*}{cccc}
\tabletypesize{\scriptsize}
\tablecaption{Best-fit QLF parameters for quasars at $z \sim 4$ \label{tbl-3}}
\tablewidth{0pt}
\tablehead{
\colhead{} & \colhead{$\beta$} & \colhead{$\Phi(M^{\ast}_{1450})$} & \colhead{$M^{\ast}_{1450}$}\\
\colhead{} & \colhead{} & \colhead{[$10^{-7}$ Mpc$^{-3}$ mag$^{-1}$]} & \colhead{[mag]}
}
\startdata
This Work & $-1.46^{+0.23}_{-0.19}$ & $3.03 \pm 0.24$ & $-24.4~(\rm{fixed})$\\
\citet{ike11} & $-1.67^{+0.11}_{-0.17}$ & $3.20 \pm 0.24$ & $-24.4 \pm 0.06$
\enddata
\tablecomments{
The bright-end slope ($\alpha$) is fixed to be $-2.58$ in both fits.
}
\end{deluxetable*}

\subsection{The QLF at $z \sim 5$}\label{subsec:4.2}

\begin{figure}
\epsscale{1.1}
\plotone{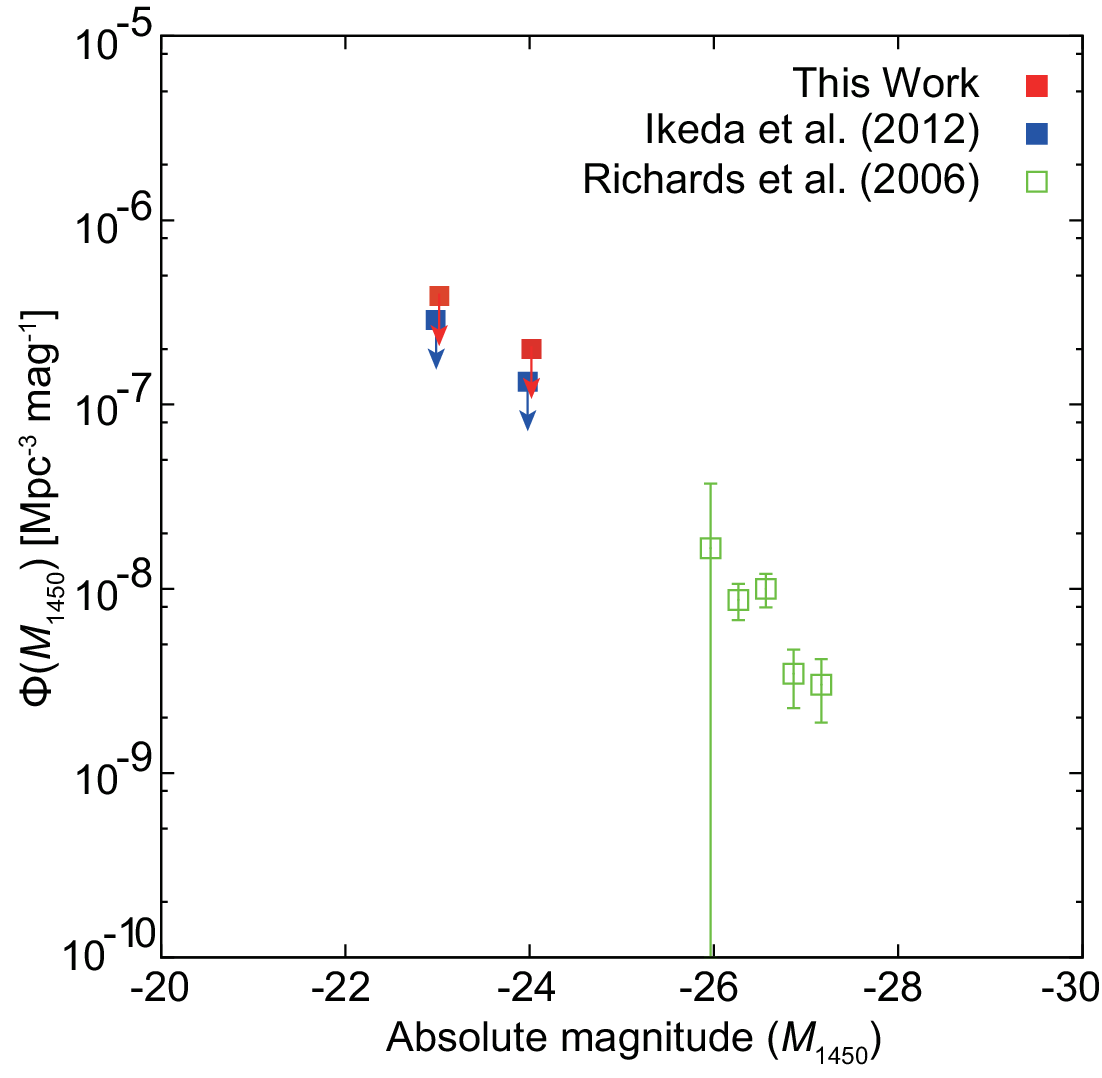}
\caption{
     Same as Figure~\ref{fig8} but for $z \sim 5$. Here the number densities of 
     COSMOS quasars are given as 1$\sigma$ upper limits. 
     The red and blue squares are the upper limits of the quasar number density 
     calculated in this work and in \citet{ike12}, respectively. 
     Green squares are the SDSS results \citep{ric06}. 
     The blue squares are slightly shifted to the left direction for clarity.
     \label{fig9}}
\end{figure}

The low-luminosity quasar survey in the COSMOS field found no 
spectroscopically-confirmed quasars at $z \sim 5$ in the magnitude range of 
$22 < i^\prime < 24$, and consequently the upper limit on the quasar number density 
was obtained \citep{ike12}. Here we investigate how the inferred upper limit is revised 
when our new completeness estimates are used. We calculate the revised 
effective comoving volume of the COSMOS quasar survey with Equation~(\ref{equ8}), 
based on the new estimates of the survey completeness given in 
Section~\ref{subsec:3.2}.
By adopting the newly derived effective comoving volume and the statistics of \citet{geh86}, 
the derived $1\sigma$ confidence upper limits on the space density of quasars are 
$\Phi < 2.00 \times 10^{-7}$ Mpc$^{-3}$ mag$^{-1}$ for $-24.52 < M_{1450} < -23.52$ and 
$\Phi < 3.89 \times 10^{-7}$ Mpc$^{-3}$ mag$^{-1}$ for $-23.52 < M_{1450} < -22.52$
(Table~\ref{tbl-4}).
Although no low-luminosity quasar was clearly identified in the FOCAS spectroscopic observation
by \citet{ike12}, here we take into account for possibilities that an object whose FOCAS spectra looks 
a type-2 AGN at $z = 5.07$ (but with a low significance) and an object without FOCAS 
spectrum could be quasars at $z \sim 5$ (see \citealt{ike12} for more details).
These results are plotted in Figure~\ref{fig9} with the previous estimates by \citet{ike12}
and also the SDSS results \citet{ric06}. 
The new results at $z \sim 5$, taking into account of the Baldwin effect and the IGM model, show $\sim$43\% 
higher number densities than the number density reported by \citet{ike12}, due to 
the smaller revised completeness at $z \sim 5$ (Section~\ref{subsec:3.2}) than the 
completeness given in \citet{ike12}.

\begin{deluxetable}{ccc}
\tabletypesize{\scriptsize}
\tablecaption{The quasar space density at $z \sim 5$ \label{tbl-4}}
\tablewidth{0pt}
\tablehead{
\colhead{} & \colhead{This Work} & \colhead{\citet{ike12}}\\
\colhead{$M_{1450}$} & \colhead{$\Phi$} & \colhead{$\Phi$}\\
 \colhead{[mag]} & \colhead{[$10^{-7}$ Mpc$^{-3}$ mag$^{-1}$]} &  \colhead{[$10^{-7}$ Mpc$^{-3}$ mag$^{-1}$]}
}
\startdata
-24.02 & $< 2.00$ & $<1.33$\\
-23.02 & $< 3.89$ & $<2.88$
\enddata
\end{deluxetable}

\subsection{The Redshift Evolution of Quasar Space Density}\label{subsec:4.3}

\begin{figure}
\epsscale{1.1}
\plotone{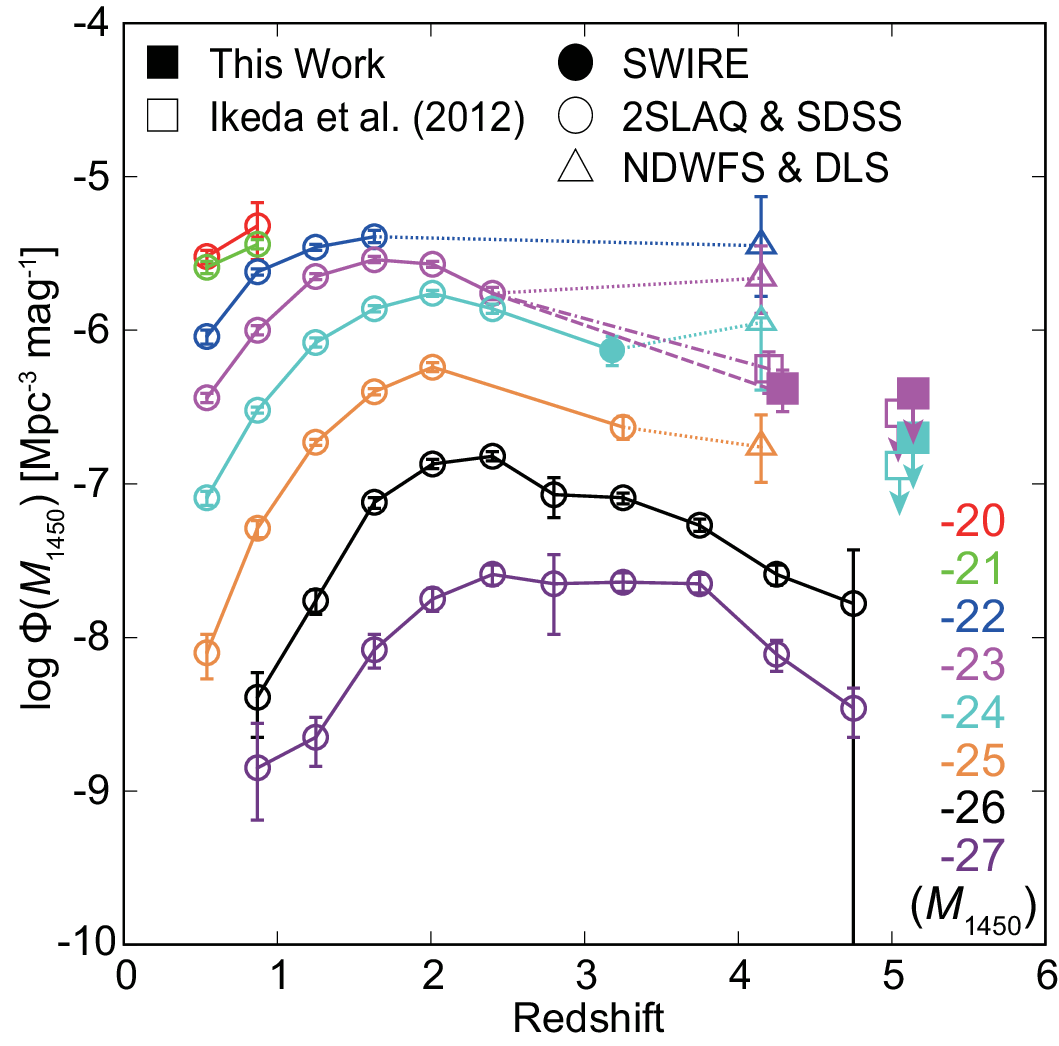}
\caption{
     Redshift evolution of the quasar space density. Red, green, blue, magenta, light 
     blue, orange, black, and purple lines are the space density of quasars with 
     $M_{1450} = -20, -21, -22, -23, -24, -25, -26, \rm {and} \ -27$, respectively.  
     Filled squares and open squares show the results of this study and those of 
     \citet{ike11, ike12}, respectively, in the COSMOS field. Open circles, filled circle, 
     and open triangles denote the results obtained in the 2dFSDSS LRG and Quasar 
     Survey \citep[2SLAQ;][]{cro09}, SDSS \citep{ric06}, the Spitzer Wide-area Infrared 
     Extragalactic Legacy Survey \citep[SWIRE;][]{sia08}, and the NOAO Deep 
     Wide-Field Survey (NDWFS) and the Deep Lens Survey (DLS) by \citet{gli11}, 
     respectively. The filled squares at $z \sim 4$ and $5$ are slightly shifted to the right direction 
     to avoid the overlap with other symbols. \label{fig10}}
\end{figure}

Here we discuss how the redshift evolution of the quasar number density depends
on the quasar luminosity (i.e., the LDDE), based on the revisited completeness and 
consequent QLF at $z \sim$ 4 and 5. As described in Section~\ref{sec:1}, the 
number density of low-luminosity quasars at high redshifts is the key to interpret the
LDDE of quasars and to discuss the cosmological evolution of SMBHs for different
masses. 
Figure~\ref{fig10} shows the quasar space density at the different absolute 
magnitude as a function of redshift. 
At $z < 4$, the quasar space density reported by the 2dFSDSS LRG and Quasar 
Survey \citep[2SLAQ;][]{cro09}, the Spitzer Wide-area Infrared Extragalactic Legacy 
Survey \citep[SWIRE;][]{sia08}, and SDSS \citep{ric06} are plotted. 
The data at $z < 4$ is totally consistent with the AGN down-sizing evolutionary
picture, as pointed out in some earlier works \citep{cro09}.
On the other hand, various results of the low-luminosity quasar space density are suggested at $z > 3$.  
The result of our study and \citet{ike11, ike12} which used the COSMOS field suggest 
that the low-luminosity quasar space density decrease from $z \sim 2$ to $z \sim 5$. This trend is suggested also by recent X-ray surveys of high-$z$ low-luminosity quasars \citep{mar16}, not only by previous X-ray surveys of relatively bright quasars \citep[e.g.,][]{ued03, has05}. However the result of the NOAO Deep Wide-Field Survey (NDWFS) 
and the Deep Lens Survey (DLS) by \citet{gli11} suggest 
constant or higher number densities of low-luminosity quasars at $z \sim 4$ than that at $z \sim 2 - 3$.  
At $z \sim 6$, recent studies \citep[e.g.,][]{wil10, kas15} suggested 
lower number density of quasar than our upper limits of them at $z \sim 5$.
Our result shows that the evolution of low-luminosity quasar number density at $z \sim 4 - 5$ 
is consistent with the AGN downsizing, even if we take into account of the Baldwin effect and the latest extinction model 
of \citet{ino14} for intergalactic absorption in the estimation of the photometric completeness.  
However, the suggested number density of low-luminosity quasars at $z \sim 4$ differs between 
the COSMOS field and the DLS/NDWFS fields. One possible reason for this discrepancy is the criterion of the point-source 
selection, that is not discussed in our work. Specifically, the DLS/NDWFS survey defines the point source based on 
ground-based $R$-band images \citep[see Figure 4 in][]{gli10}, while the COSMOS survey defines the point source based on 
the HST images \citep{ike11}. Another possible reason for the discrepancy in the number density of high-$z$ low-luminosity quasars 
in previous surveys is the small sample size of their spectroscopically-confirmed quasars. To clarify the reason for
this difference and understand the evolution of the quasar number density more accurately, 
further observations of low-luminosity quasars in a wider survey area are crucial.

As shown in Sections~\ref{subsec:4.1} and \ref{subsec:4.2}, the inferred number 
density of high-redshift low-luminosity quasars was overestimated or underestimated
by a factor of $\sim20-50$\% compared with the result of \citet{ike11, ike12}, depending on their redshift and luminosity. This
suggests that there was a systematic error caused by the adopted IGM extinction 
model and the luminosity dependence of quasar spectral template (i.e., the Baldwin
effect) in the quasar number density reported in the previous surveys. Note that
this systematic error is not seriously large if it is compared with the statistical error
in the COSMOS works \citep{ike11, ike12}, because of the small number of
photometric and spectroscopic sample of high-$z$ low-luminosity quasars. However,
near-future surveys of high-$z$ low-luminosity quasars such as Subaru/Hyper
Suprime-Cam (\citealt{miy12} , see also \citealt{maty16}) and Subaru/Prime Focus
Spectrograph \citep{tak14} will observe a numerous number of high-$z$ low-luminosity
quasars. In those surveys, systematic errors with a factor of $\sim20-50$\% is not
negligible at all given their small statistical error. Therefore, understanding the
proper method to infer the survey completeness and effective volume as investigated
in this work is crucial.

\section{SUMMARY}\label{sec:5}

In order to derive the high-$z$ low-luminosity QLF accurately, we quantify the influence of
the Baldwin effect and the IGM absorption model on the quasar color. We then
revaluate the photometric completeness of low-luminosity quasar surveys with 
COSMOS images.  The main results of this study are as follows.

\begin{enumerate}
    \item \ \ At $z \sim 4 - 5$, lower-luminosity quasars are more easily selected by color selection criteria than higher-luminosity quasars when the effects of the Baldwin effect are properly considered.
    \item \ \ By comparing the quasar model spectra adopting the extinction model by \citet{ino14} with the other model by \citet{mad95}, we find that the former (latest) model predicts a smaller Lyman-break feature at $z \sim 4 - 5$, that makes the color selection of quasars more challenging. 
    \item \ \ The revised completeness is larger at $z \sim 4$ but smaller at $z \sim 5$ than our previous works \citep{ike11, ike12}, due to the effects of the Baldwin effect and IGM model.
    \item \ \ The revaluated quasar space densities at $z \sim 4$ are $(3.49 \pm 1.62) \times 10^{-7}$ Mpc$^{-3}$ mag$^{-1}$ 
for $-24.09 < M_{1450} < -23.09$ and $(5.24 \pm 2.13) \times 10^{-7}$ Mpc$^{-3}$ mag$^{-1}$ 
for $-23.09 < M_{1450} < -22.09$, respectively. These densities are lower than the results without considering the Baldwin effect \citep{ike11}.  
Therefore the revaluated faint-end slope of our QLF, $\beta$ $= -1.46^{+0.23}_{-0.19}$, is flatter than that of \citet{ike11}.
    \item \ \ The upper limits of the quasar space density at $z \sim 5$ are $\Phi < 2.00 \times 10^{-7}$ Mpc$^{-3}$ mag$^{-1}$ 
for $-24.52 < M_{1450} < -23.52$ and $\Phi < 3.89 \times 10^{-7}$ Mpc$^{-3}$ mag$^{-1}$ 
for $-23.52 < M_{1450} < -22.52$. These densities are higher than the results without considering the Baldwin effect.
    \item \ \ Even by taking account of the revisions in the completeness estimate, the inferred luminosity-dependent density evolution of quasars is consistent with the AGN down-sizing evolutionary picture. 
\end{enumerate}

\acknowledgments 
We would like to thank Masaru Kajisawa, Masayuki Akiyama, and Kazuyuki Ogura for their helpful 
comments and suggestions. This work was financially supported in part by JSPS 
(TN: 25707010, 16H01101, and 16H03958; YT: 23244031 and 16H02166) and also by the 
JGC-S Scholarship Foundation. We thank the Subaru staff for their invaluable help for obtaining the COSMOS 
imaging data and spectroscopic follow-up data, and also all members of the COSMOS 
team. We would also like to thank an anonymous referee for her/his
useful comments and suggestions. Data analysis were in part carried out on common use data analysis computer system 
at the Astronomy Data Center, ADC, of the National Astronomical Observatory of Japan.
Funding for SDSS-III has been provided by the Alfred P. Sloan Foundation, the 
Participating Institutions, the National Science Foundation, and the U.S. Department 
of Energy Office of Science. The SDSS-III web site is \url{http://www.sdss3.org/}.
SDSS-III is managed by the Astrophysical Research Consortium for the Participating 
Institutions of the SDSS-III Collaboration including the University of Arizona, the 
Brazilian Participation Group, Brookhaven National Laboratory, Carnegie Mellon 
University, University of Florida, the French Participation Group, the German 
Participation Group, Harvard University, the Instituto de Astrofisica de Canarias, the 
Michigan State/Notre Dame/JINA Participation Group, Johns Hopkins University, 
Lawrence Berkeley National Laboratory, Max Planck Institute for Astrophysics, Max 
Planck Institute for Extraterrestrial Physics, New Mexico State University, New York 
University, Ohio State University, Pennsylvania State University, University of 
Portsmouth, Princeton University, the Spanish Participation Group, University of 
Tokyo, University of Utah, Vanderbilt University, University of Virginia, University of 
Washington, and Yale University. 
IRAF is distributed by the National Optical Astronomy Observatory, which is operated 
by the Association of Universities for Research in Astronomy (AURA) under a 
cooperative agreement with the National Science Foundation.

\appendix

\section{Dependence of $EW$({C~{\sc iv}}) on the quasar luminosity and redshift}\label{app}

\begin{figure*}
\epsscale{1.15}
\plotone{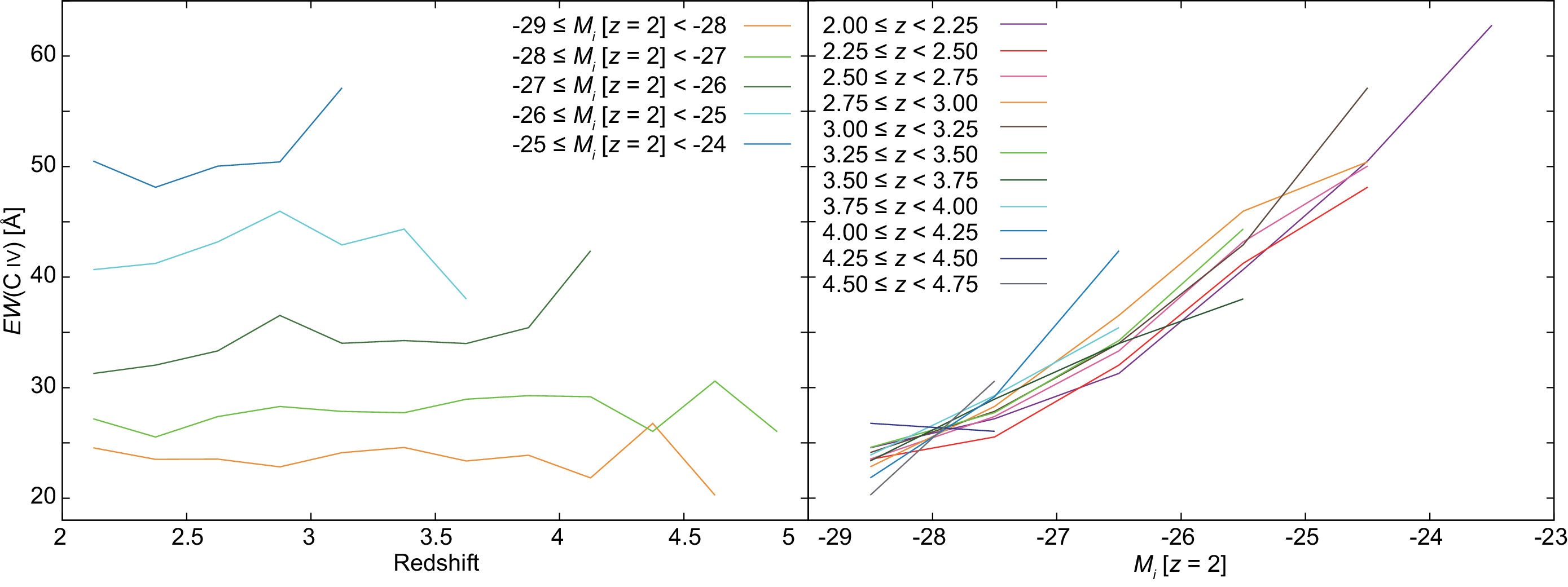}
\caption{The relation between the rest-frame $EW$({C~{\sc iv}}) of composite spectra and redshift (left panel) and luminosity (right panel). Only the data made with $> 20$ quasars are plotted. The line color corresponds to the difference of luminosity (left panel) and difference of redshift (right panel), respectively.  
\label{fig11}}
\end{figure*}

In Section~\ref{subsec:2.2}, we described that $EW$({C~{\sc iv}}) of the quasar spectrum does not depend on redshift. 
To clarify this statement, we make composite spectra for each luminosity and redshift bin.
We first divide the $43,953$ quasar spectra selected from the BOSS quasar catalog 
for each  $\Delta M_i[z=2] = 1$ and  $\Delta z = 0.25$ bin.  
The number of quasars in each bin is showed in Table~\ref{tbl-5}.  
We make composite spectra for each bin and measure $EW$({C~{\sc iv}}) of them in the rest frame.
The measured $EW$({C~{\sc iv}}) is summarized in Table~\ref{tbl-6}.
In Figure~\ref{fig11}, we show the measured $EW$({C~{\sc iv}}) as a function of redshift (left panel) and luminosity (right panel), respectively.
This figure suggests that $EW$({C~{\sc iv}}) of the quasar spectrum correlates significantly with the luminosity 
but does not correlate with the redshift.

\begin{deluxetable}{cccccccccccccc}
\tabletypesize{\scriptsize}
\tablecaption{The number of quasar spectra for making composite spectra\label{tbl-5}}
\tablewidth{0pt}
\tablehead{
\colhead{}& 
\colhead{$2.00 \le z < 2.25$} & \colhead{$2.25 \le z < 2.50$} & 
\colhead{$2.50 \le z < 2.75$} & \colhead{$2.75 \le z < 3.00$} &
\colhead{$3.00 \le z < 3.25$} & \colhead{$3.25 \le z < 3.50$} &
}
\startdata
$-31 \le M_i\tablenotemark{a} < -30$ & 0 & 0 & 0 & 4 & 1 & 0   \\
$-30 \le M_i < -29$ & 7 & 7 & 6 & 13 & 7 & 10   \\
$-29 \le M_i < -28$ & 94 & 154 & 132 & 124 & 129 & 115   \\
$-28 \le M_i < -27$ & 438 & 1181 & 978 & 727 & 738 & 526   \\
$-27 \le M_i < -26$ & 1655 & 4421 & 3060 & 1895 & 1713 & 833   \\
$-26 \le M_i < -25$ & 3344 & 7230 & 3964 & 1852 & 1003 & 266   \\
$-25 \le M_i < -24$ & 1651 & 2534 & 746 & 174 & 25 & 1   \\
$-24 \le M_i < -23$ & 44 & 16 & 1 & 1 & 0 & 1   \\ \hline
$$ &
$3.50 \le z < 3.75$ & $3.75 \le z < 4.00$ &
$4.00 \le z < 4.25$ & $4.25 \le z < 4.50$ &
$4.50 \le z < 4.75$ & $4.75 \le z < 5.00$ \\ \hline
$-31 \le M_i < -30$ & 0 & 0 & 0 & 0 & 0 & 0   \\
$-30 \le M_i < -29$ & 5 & 9 & 9 & 8 & 2 & 1   \\
$-29 \le M_i < -28$ & 91 & 81 & 57 & 34 & 23 & 13   \\
$-28 \le M_i < -27$ & 350 & 296 & 132 & 57 & 34 & 22   \\
$-27 \le M_i < -26$ & 466 & 282 & 75 & 18 & 4 & 2   \\
$-26 \le M_i < -25$ & 46 & 13 & 1 & 0 & 0 & 0   \\
$-25 \le M_i < -24$ & 0 & 0 & 0 & 0 & 0 & 0   \\
$-24 \le M_i < -23$ & 1 & 0 & 0 & 0 & 0 & 0
\enddata
\tablenotetext{a}{$M_i$ denotes $M_i [z = 2]$.}
\end{deluxetable}

\begin{deluxetable}{cccccccccccccc}
\tabletypesize{\scriptsize}
\tablecaption{The $EW$({C~{\sc iv}}) [$\rm{\AA}$] of composite spectra at rest frame\label{tbl-6}}
\tablewidth{0pt}
\tablehead{
\colhead{}&
\colhead{$2.00 \le z < 2.25$} & \colhead{$2.25 \le z < 2.50$} & 
\colhead{$2.50 \le z < 2.75$} & \colhead{$2.75 \le z < 3.00$} &
\colhead{$3.00 \le z < 3.25$} & \colhead{$3.25 \le z < 3.50$} &
}
\startdata
$-31 \le M_i\tablenotemark{a} < -30$ & \nodata & \nodata & \nodata & 13.3 & 29.8 & \nodata   \\
$-30 \le M_i < -29$ & 23.4 & 19.5 & 14.3 & 23.2 & 18.9 & 23.8   \\
$-29 \le M_i < -28$ & 24.6 & 23.5 & 23.5 & 22.8 & 24.1 & 24.6   \\
$-28 \le M_i < -27$ & 27.2 & 25.5 & 27.4 & 28.3 & 27.9 & 27.7   \\
$-27 \le M_i < -26$ & 31.3 & 32.0 & 33.3 & 36.5 & 34.0 & 34.3   \\
$-26 \le M_i < -25$ & 40.7 & 41.2 & 43.2 & 46.0 & 42.9 & 44.3   \\
$-25 \le M_i < -24$ & 50.5 & 48.1 & 50.0 & 50.4 & 57.1 & 56.6   \\
$-24 \le M_i < -23$ & 62.8 & 57.5 & 48.6 & 21.1 & \nodata & 29.1   \\ \hline
$$ &
$3.50 \le z < 3.75$ & $3.75 \le z < 4.00$ &
$4.00 \le z < 4.25$ & $4.25 \le z < 4.50$ &
$4.50 \le z < 4.75$ & $4.75 \le z < 5.00$ \\ \hline
$-31 \le M_i < -30$ & \nodata & \nodata & \nodata & \nodata & \nodata & \nodata   \\
$-30 \le M_i < -29$ & 20.9 & 29.2 & 21.8 & 20.2 & 19.9 & 24.3   \\
$-29 \le M_i < -28$ & 23.4 & 23.9 & 21.8 & 26.8 & 20.3 & 21.2   \\
$-28 \le M_i < -27$ & 29.0 & 29.3 & 29.2 & 26.0 & 30.6 & 26.0   \\
$-27 \le M_i < -26$ & 34.0 & 35.4 & 42.4 & 36.2 & 15.4 & 50.3   \\
$-26 \le M_i < -25$ & 38.0 & 26.9 & 82.0 & \nodata & \nodata & \nodata   \\
$-25 \le M_i < -24$ & \nodata & \nodata & \nodata & \nodata & \nodata & \nodata   \\
$-24 \le M_i < -23$ & 17.3 & \nodata & \nodata & \nodata & \nodata & \nodata
\enddata
\tablecomments{
For the case if a bin contains only one quasar, the $EW$({C~{\sc iv}}) measured in the individual spectrum of the corresponding quasar is given.
}
\tablenotetext{a}{$M_i$ denotes $M_i [z = 2]$.}
\end{deluxetable}


\end{document}